\documentclass[a4paper,10pt]{article}

\usepackage[utf8]{inputenc}
\usepackage[T1]{fontenc}

\usepackage{a4wide}

\usepackage{amsmath,amssymb}
\usepackage{amsthm}
\usepackage{enumerate} 

\usepackage{authblk}

\usepackage[dvipsnames,table,xcdraw,svgnames]{xcolor}
\definecolor{highlightNEW}{named}{black}

\usepackage{hyperref}
\hypersetup{
    colorlinks=true,
    linkcolor=Navy,
    citecolor=Navy,
    urlcolor=Navy
}

\usepackage{graphicx}
\graphicspath{{./fig/},{./fig-thesis/},{./fig-thesis/varAnalysis/},{./fig-thesis/calibration/},{./fig-thesis/bootcalibrations/ScatVar/},{./fig-thesis/bootcalibrations/ScatExp/},{./fig-thesis/bootcalibrations/ScatMatrix/}}
\usepackage{epstopdf}
\usepackage{subcaption}
\usepackage[section]{placeins} 

\usepackage{booktabs}
\usepackage{multirow}

\usepackage[absolute,overlay,showboxes]{textpos}
\TPMargin{2mm}
\textblockrulecolour{Navy}

\usepackage[round]{natbib}
\bibpunct[, ]{(}{)}{;}{a}{}{,}

\newtheorem{theorem}{Theorem}[section] 
 
\newtheorem{example}[theorem]{Example}

\newcommand{\doi}[1]{DOI~\href{\detokenize{https://dx.doi.org/#1}}{\detokenize{#1}}}

\newdimen\CdotAxis
\newcommand*{\CdotAux}[3]{%
  {%
    \settoheight\CdotAxis{$#2\vcenter{}$}%
    \sbox0{%
      \raisebox\CdotAxis{%
        \scalebox{#1}{%
          \raisebox{-\CdotAxis}{%
            $\mathsurround=0pt #2#3$%
          }%
        }%
      }%
    }%
    \dp0=0pt %
    \sbox2{$#2\bullet$}%
    \ifdim\ht2<\ht0 %
      \ht0=\ht2 %
    \fi
    \sbox2{$\mathsurround=0pt #2#3$}%
    \hbox to \wd2{\hss\usebox{0}\hss}%
  }%
}
\makeatletter
\def\mathcolor#1#{\@mathcolor{#1}}
\def\@mathcolor#1#2#3{%
  \protect\leavevmode
  \begingroup
    \color#1{#2}#3%
  \endgroup
}
\makeatother

\let\oldalpha\alpha
\renewcommand{\alpha}{\mathcolor{highlightNEW}{\oldalpha}}
\newcommand{\ccode}[2]{\par
        \vspace*{8pt}
        {{\leftskip18pt\rightskip\leftskip
        \noindent{\it #1}\/: #2\par}}\par}
\newcommand{\keywords}[1]{\ccode{Keywords}{#1}}
\newcommand{\email}[1]{\href{mailto:#1}{#1}}

\def\received#1{Received~#1\par}
\def\revised#1{Revised~#1\par}
\def\accepted#1{Accepted~#1\par}
\def\published#1{Published~#1\par}

\def\foliofont{\fontsize{8}{10}\selectfont}
\usepackage[mathscr]{euscript}
\DeclareSymbolFont{rsfs}{U}{rsfs}{m}{n}
\DeclareSymbolFontAlphabet{\mathscrsfs}{rsfs}

\newcommand{\jpTitle}{Isogeometric analysis in option pricing}
\newcommand{\jpAuthors}{J. Posp\'{\i}\v{s}il and V. \v{S}v\'{\i}gler}
\newcommand{\jpKeywords}{isogeometric analysis; option pricing; NURBS; finite element method; stochastic volatility models}
\newcommand{\jpMSC}{35R09; 65M60; 65D07; 91G20; 91G6}
\newcommand{\jpJEL}{C58; C63; G12}
\newcommand{\jpDateReceived}{23 October 2017} 
\newcommand{\jpDateRevised}{Revised 5 April 2018}
\newcommand{\jpDateAccepted}{8 June 2018}
\newcommand{\jpDatePublished}{16 July 2018}
\newcommand{\jpDate}{}

\author[1]{Jan Posp\'{\i}\v{s}il\thanks{Corresponding author, \email{honik@kma.zcu.cz}}} 
\author[1]{Vladim\'{\i}r \v{S}v\'{\i}gler} 
\affil[1]{NTIS - New Technologies for the Information Society, Faculty of Applied Sciences, \authorcr University of West Bohemia, Univerzitn\'{\i} 2732/8, 301 00 Plze\v{n}, Czech Republic,\vspace*{3pt}}

\ifpdf
\hypersetup{
  pdftitle={\jpTitle},
  pdfauthor={\jpAuthors},
  pdfkeywords={\jpKeywords},
  pdfinfo={
      MSC={\jpMSC},
      JEL={\jpJEL}
  }
}
\fi

\title{\textcolor{Navy}{\textsc{\jpTitle}}}
\date{\jpDate}

\usepackage[absolute,overlay,showboxes]{textpos}
\TPMargin{2mm}
\textblockrulecolour{Navy}

\begin{document}

\maketitle
\begin{textblock}{6}(1.25,1.15)
{\foliofont\noindent This is an Accepted Manuscript of an article published by Taylor \& Francis in 
International Journal of Computer Mathematics 
96(11), 2177--2200, 2019, \doi{10.1080/00207160.2018.1494826}. \\
Available online \url{https://www.tandfonline.com/10.1080/00207160.2018.1494826}.
}
\end{textblock}

\begin{center}
\received{\jpDateReceived}
\revised{\jpDateRevised}
\accepted{\jpDateAccepted}
\published{\jpDatePublished}
\end{center}

\begin{abstract}
Isogeometric analysis is a recently developed computational approach that integrates finite element analysis directly into design described by non-uniform rational B-splines (NURBS). In this paper we show that price surfaces that occur in option pricing can be easily described by NURBS surfaces. For a class of stochastic volatility models, we develop a methodology for solving corresponding pricing partial integro-differential equations numerically by isogeometric analysis tools and show that a very small number of space discretization steps can be used to obtain sufficiently accurate results. Presented solution by finite element method is especially useful for practitioners dealing with derivatives where closed-form solution is not available.

\end{abstract}

\keywords{\jpKeywords}
\ccode{MSC classification}{\jpMSC}
\ccode{JEL classification}{\jpJEL}

\section{Introduction}\label{sec:intro}

Isogeometric analysis is a computational approach that integrates finite element analysis directly into design described by non-uniform rational B-splines (NURBS). Although the ideas of having spline-based finite elements method (FEM) goes back to the early stages of FEMs development, only a recent development of computer systems gave birth to this new field that is widely accepted especially in computational mechanics and computer-aided geometrical modelling. These two historically separate disciplines have joined forces and embraced the vision of isogeometric analysis, in order to simplify the model development process by integrating engineering design and analysis into a unified framework. This allows models to be designed, tested and adjusted in one go, using a common data set. We aim to show, how these ideas can be applied in finance. 

It is widely accepted that the isogeometric analysis began with the publication of the paper by \cite{Hughes05}. Since then, and especially after publishing the book by \cite{Cottrell09}, isogeometric analysis attracted considerable attention in the research community, and at the present time it is enjoying exponential growth, as measured by the number of papers published on the topic and the citations to them in the literature. It is not the aim of this paper to give a detailed review of what has been done in isogeometric analysis recently, we refer the reader especially to the series of annual International Conferences on Isogeometric Analysis with selected papers being published in special issues of the journal Computer Methods in Applied Mechanics and Engineering.

The aim and novelty of this manuscript lies in application of isogeometric analysis in mathematical finance, namely in option pricing. Although the rectangular domains considered in option pricing equations are rather simple from the geometrical point of view, price surfaces obtained as a solution to these equations are on the other hand quite complex. We show that these surfaces can be easily described by NURBS surfaces. We develop a methodology for solving corresponding pricing partial integro-differential equations (PIDEs) numerically by isogeometric analysis tools, i.e. by FEM with NURBS basis functions. We compare the method and the results to closed-form solutions where available. Although NURBS are computationally more demanding than standard basis functions, we show that a very small number of space discretization steps can be used to obtain sufficiently good results. The FEM results are essential especially for practitioners dealing with derivatives where closed-form solution is not available.

In this paper we study the constant volatility jump diffusion model by \cite{Merton76} (and the \cite{BlackScholes73} model as a special case) and approximative fractional stochastic volatility jump diffusion model recently proposed by \cite{PospisilSobotka16amf} (and the models by \cite{Bates96} and \cite{Heston93} as special cases). For European style options with path-independent payoffs, these models offer a semi-closed pricing formula. Motivation for pricing options using a numerical solution of the corresponding PIDE are of course exotic derivative securities with American payoff style whose semi-closed pricing formulas are not available. The problem of pricing American options leads to the problem of solving of variational inequalities. This paper gives a fundamental framework for the analytical and numerical setting of the problem. Although the proposed methodology is designed in order to allow further extensions, pricing American options goes beyond the aims of this manuscript. 

Solving PIDEs is a challenging research topic not only in mathematical finance, but in theoretical mathematical and numerical analysis. In finance, most of the publications focus on solving the PIDEs arising from the Merton model. The two main numerical approaches are finding the solution with the finite elements or finite differences methods. Finite differences methods are studied, for example, by \cite{SalmiToivanen14,Fakharany16,Hout16}. A wide class of models is analysed by \cite{Fakharany16} where the pricing of European and American options was performed. Also, various L\'{e}vy measures were used (Kobol, Meixner and generalized hyperbolic) therein. The properties of implicit-explicit scheme were studied in \cite{SalmiToivanen14} together with Fourier stability analysis of the method. Further improvement in high-order splitting schemes for forward and backward PDEs and PIDEs arising in option pricing is presented in \cite{Itkin15}. Splitting schemes of the Alternating Direction Implicit (ADI) type are used by \cite{Hout16} to price both European and American put options under the Merton or Bates model. 

Lately, the regime-switching jump diffusion processes were studied, where the parameters determining the behaviour of the jump process switch between various regimes, see e.g. \cite{Dang16}, \cite{RambeerichPantelous16}. The solvability of PIDE by \cite{Dang16} is proved by constructing a sequence of sub-solutions which also define an effective numerical algorithm. The method of finite elements is used in \cite{RambeerichPantelous16} where European, American and Butterfly options are priced. 

Papers studying partial differential equations in models with non-constant volatility usually assume a local volatility. This is true especially for articles about American options and articles about jump diffusion models \citep{ContVoltchkova05a,TankovVoltchkova09}. Stochastic volatility (SV) models are nicely presented in the book by \cite{Fouque00}. Typically, to make the numerical solution of SV models well-posed, conditions on the parameters (like correlation) must be laid as was shown by \cite{Lions07}. Pricing American options under SV models is studied for example by \cite{AitSahlia10a} with the empirical results in \cite{AitSahlia10b}. The numerical influence of the stochastic volatility on the foreign equity option prices is analysed by \cite{SunXu15}. Further studies have shown the importance of stochastic volatility jump diffusion (SVJD) models e.g. \cite{Sun15} and numerical properties of the solutions of corresponding PIDEs, e.g. \cite{Aboulaich13a}. 

Other numerical methods such as quadratic spline collocation \citep{ChristaraLeung16}, adaptive wavelet collocation method \citep{Li14} or a mixed PDE and Monte Carlo method \citep{Loeper09,Lipp13} were also studied. A wide class of pricing methods were tested in the BENCHOP project \citep{Sydow15}, where 15 different numerical methods were compared for 6 benchmark problems. Apart from the Monte Carlo and finite differences methods, a special attention was paid to Fourier methods and radial basis functions (RBF) methods. Although Fourier methods rely on the availability of the characteristic function of the underlying stochastic process, they can provide a reasonably good solution very quickly, especially if the fast Fourier transform method is used \citep{CarrMadan99,Lord08} or the so-called COS method \citep{Fang08} that uses Fourier cosine series expansions. Since the original paper by \cite{Hon99} presenting the application of RBF to solve the option pricing PDEs, many papers studying different basis functions or different node locations were published. However, only recently, RBF method for Merton model was used by \cite{Chan16} and only little is known for using RBF method for PIDEs in SVJD models.

To conclude the brief literature review we have to say that there are not many publications about solving PIDEs using FEM. We refer the reader to the review paper about variational methods in derivative pricing written by \cite{Feng07}.

Our paper is structured as follows. In Section \ref{sec:preliminaries}, we introduce the B-spline and NURBS basis functions and curves. In particular we show, how B-Splines and NURBS can be fitted to smooth, non-smooth or even discontinuous functions easily. We also introduce the considered option pricing models, in particular, the constant volatility jump diffusion model by \cite{Merton76} (and the \cite{BlackScholes73} model as a special case), stochastic volatility model by \cite{Bates96} (and the \cite{Heston93} model as a special case) and last but not least a recently proposed approximative fractional stochastic volatility jump diffusion model \citep{PospisilSobotka16amf, BaustianMrazekPospisilSobotka17asmb}.

In Section \ref{sec:methodology}, we derive the variation formulation of studied PIDEs and show, how to solve the pricing equations numerically by isogeometric analysis tools, i.e. by FEM with NURBS elements.

In Section \ref{sec:results}, we show the results of fitting NURBS to exact pricing formulas. We also provide the results of the FEM solutions that we also compare to the closed-form solutions.
We conclude in Section \ref{sec:conclusion}.

\section{Preliminaries}\label{sec:preliminaries}

\subsection{NURBS basis functions and curves}

\emph{B-spline basis functions} are piecewise polynomial smooth functions defined by a recursive scheme. 
A \emph{knot vector} $C = (c_1, \ldots, c_{m})^T$ is a nondecreasing vector of $m$ real-valued coordinates in the parameter space such that 
\begin{equation}
m = n + p + 1,
\end{equation}
where $n$ is the number of basis functions used to construct the B-spline curve and $p$ is the polynomial order. Knots partition the parameter space into elements either \emph{uniformly}, if they are equally spaced in the parameter space, or we say that the vector is \emph{non-uniform}. Knot values may be repeated, i.e. more than one knot may have the same value. Multiplicities of knots have important implications for the properties of the basis. A knot vector is said to be \emph{open} if its first and last knot values are repeated $p+1$ times.

The B-spline basis of the degree zero ($p=0$) is defined as a piecewise constant
\begin{equation}
N_{i,0}(\xi) = 
\begin{cases}
1, & c_i \leq \xi < c_{i+1}, \\
0, & \text{otherwise},
\end{cases}
\end{equation}
for $i = 1,\ldots, n$. The higher order B-spline basis functions are defined recursively as
\begin{equation}
N_{i,p}(\xi) = \frac{\xi-c_{i}}{c_{i+p}-c_{i}} N_{i,p-1}(\xi) + \frac{c_{i+p+1}-\xi}{c_{i+p+1}-c_{i+1}} N_{i+1,p-1}(\xi), \\
\end{equation}
for $i=1,2,\ldots,n$ and $p=1,2,3,\dots$. In case of repeated knots, some denominators in the recurrent definition can be zero, if this happens, the whole fraction is defined to be zero.  
From now on, we assume the choice of the knot vector $C$ and the degree $p$ such that no basis function $N_{i,p}(\xi)$ is identically zero. 
The list of some properties of B-spline basis follows: 
\begin{enumerate}
\item the basis functions $N_{i,p}(\xi)$ are all piecewise polynomial, 
\item the sum of all basis functions $\sum_{i=1}^{n} N_{i,p}(\xi)$ for $\xi\in[c_1,c_{m}]$ is equal to a function being identically equal to one, 
\item all basis functions are nonnegative, i.e. $N_{i,p}(\xi)\geq 0$ for all $\xi$.
\end{enumerate} 
Derivatives $N'_{i,p}(\xi) = \frac{\mathrm{d}}{\mathrm{d}\xi} N_{i,p}(\xi)$ of the B-spline basis functions can be easily computed alongside with the original basis functions as
\begin{align}
N'_{i,0}(\xi) &= 0, \\
N'_{i,p}(\xi) &= \frac{p}{c_{i+p}-c_i} N_{i,p-1}(\xi) - \frac{p}{c_{i+p+1}-c_{i+1}} N_{i+1,p-1}(\xi),
\end{align}
for $i=1,2, \ldots, n$ and $p=1,2,\dots$. 

A \emph{B-spline curve} in $\mathbb{R}^d$ is defined as a linear combination of B-spline basis functions, the vector-valued coefficients are referred to as \emph{control points}. Given $n$ basis functions $N_{i,p}(\xi), i=1,2,\dots,n$ and corresponding control points $\mathbf{P}_i\in\mathbb{R}^d, i=1,2,\dots,n$, a piecewise polynomial B-spline curve of order $p$ is given by
\begin{equation}\label{eq:B-spline-curve}
\mathbf{C}(\xi) = \sum\limits_{i=1}^n N_{i,p}(\xi) \mathbf{P}_i.
\end{equation}

Let $\mathbf{w} = (w_1, w_2, \ldots, w_{n})^T$ be a \emph{weight vector} such that $w_i>0$ for $i=1,2,\ldots, n$. 
Then, we can define \emph{NURBS basis functions} by 
\begin{equation}\label{eq:NURBS-basis}
R_{i}^p(\xi) 
= \frac{w_i N_{i,p}(\xi)}{\sum_{j=1}^{n} w_j N_{j,p}(\xi)}.
\end{equation}
Note that the expression in the denominator can be simplified for computational efficiency -- only the parts of the functions $N_{j,p}(\xi)$ whose nonzero part coincides with the nonzero part of the function $N_{j,p}(\xi)$ need to be summed. Thus, we can list some properties of the NURBS basis functions:
\begin{enumerate}
\item the basis functions $R_i^p(\xi)$ are piecewise rational; since it is defined as a ratio of two piecewise polynomials of order $p$, it is also often referred to as having order $p$ and common names \emph{quadratic} ($p=2$) basis function, \emph{cubic} ($p=3$) and similar are often used in this sense, 
\item the sum of all basis functions $\sum_{i=1}^n R_i^p(\xi)$ is equal to a function being identically equal to one, 
\item all the basis functions $R_i^p(\xi)$ are nonnegative, 
\item every basis function $R_{i}^p(\xi)$ has the same support as the corresponding $N_{i,p}(\xi)$.
\end{enumerate}
The derivatives of the NURBS basis functions are 
\begin{align}
\frac{\mathrm{d}}{\mathrm{d}\xi} R_{i}^p(\xi) 
&= \frac{w_i N'_{i,p}(\xi)}{\sum_{j=1}^{n} w_j N_{j,p}(\xi)} - R_i^p(\xi) \frac{\sum_{j=1}^{n} w_j N'_{j,p}(\xi)}{\sum_{j=1}^{n} w_j N_{j,p}(\xi)}.
\end{align}
A \emph{NURBS curve} defined for the same control points as in \eqref{eq:B-spline-curve} is defined as
\begin{equation}\label{eq:NURBS-curve}
\mathbf{C}(\xi) = \sum\limits_{i=1}^n R_i^p(\xi) \mathbf{P}_i.
\end{equation}

\begin{figure}[ht!]
\includegraphics[width=\textwidth]{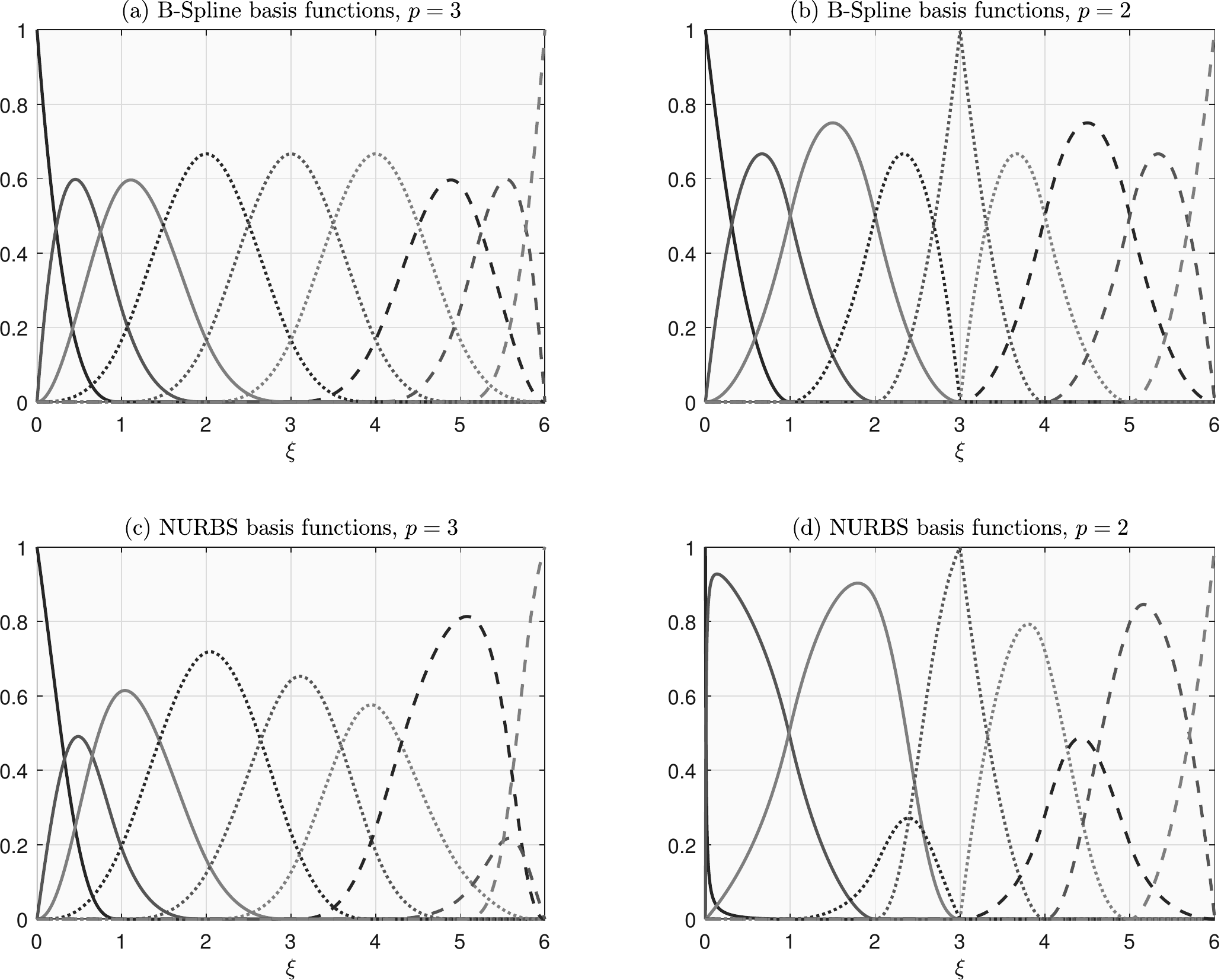}
\caption{
\textbf{(a)} A cubic ($p=3$) B-spline basis with open knot vector $C = (0,0,0,0,1,2,3,4,5,6,6,6,6)$. 
\textbf{(b)} A quadratic ($p=2$) B-spline basis with open, non-uniform knot vector $C = (0,0,0,1,2,3,3,4,5,6,6,6)$ (note that the point $3$ being present 2 times creates nonsmoothness at it). 
\textbf{(c)} A cubic NURBS basis constructed from the basis (a) with weight vector $w = (0.87, 0.48, 0.66, 0.74, 0.51, 0.4, 0.68, 0.11, 0.5)$, 
\textbf{(d)} A quadratic NURBS basis constructed from the basis (b) with weight vector $w = (0.01, 0.81, 0.86, 0.14, 0.58, 0.54, 0.21, 0.83, 0.78)$. } 
\label{fig:nurbs-basis}
\end{figure}

In Figure \ref{fig:nurbs-basis}, we can see examples of various bases plotted in the parameter space $[0,6]$, i.e. all $c_i\in[0,6], i=1,2,\dots,m$. In all cases $n=9$ and either $m=12$ for the second degree (quadratic) basis functions or $m=13$ for the third degree (cubic) basis functions. Multiplicity of knots can be easily used, for example, to describe also non-smooth "peaks". We take advantage of this fact later in describing the non-smooth payoff functions arising in the initial (or in fact terminal) conditions for pricing differential equations. Rationality of NURBS then gives us much greater flexibility (compared to the standard B-splines) in describing complicated solutions of these equations. It is worth to realize that non-smooth payoff functions that are piecewise polynomial of order up to $p$ can be described by NURBS of order $p$ \emph{exactly}. For the same reason, piecewise linear basis functions (that have a value of 1 at their respective nodes and 0 at other nodes) are special case of NURBS basis functions. Many useful properties and geometric algorithms for NURBS curves and surfaces can be found in the famous NURBS book by \cite{PieglTiller2012}, especially in Chapters 5 and 6.

To demonstrate the above-mentioned properties we show how to fit NURBS to several functions of interest. For the sake of simplicity, let us  perform the fit in the parametric space, i.e. given a knot vector $C$ and the order $p$, we are looking for the best fit to the function $f(\xi)$, $\xi\in[c_1,c_m]$ either by a B-spline -- a linear combination of $n = m - p -1$ B-spline basis functions $N_{i,p}(\xi)$, 
\[ \hat{f}_{\text{bs}}(\xi) = \sum_{i=1}^n \hat{a}_i N_{i,p}(\xi), \]
where coefficients $\hat{a}_i$ are to be determined or by a NURBS
\[ \hat{f}_{\text{nrb}}(\xi) = \sum_{i=1}^n \hat{b}_i R_i^p(\xi), \]
where $\hat{b}_i$ and the weights $w_i>0$, $i=1,2,\dots,n$ are also to be determined. The best fit will be measured by the mean $L^2$ errors 
\[ 
\hat{\epsilon}_{\text{bs}} = \frac1{c_m-c_1} \int_{c_1}^{c_m} (f(\xi)-\hat{f}_{\text{bs}}(\xi))^2\,\mathrm{d}\xi
\quad\text{and}\quad
\hat{\epsilon}_{\text{nrb}} = \frac1{c_m-c_1} \int_{c_1}^{c_m} (f(\xi)-\hat{f}_{\text{nrb}}(\xi))^2\,\mathrm{d}\xi, \]
where $\hat{\epsilon}_{\text{nrb}}$ is minimized with respect to the weights $w_i$.

\begin{example}[NURBS fit to a smooth function]\label{ex:nurbs_fit_exp}
It is clear that polynomials of order $p$ can be fitted both by B-splines and NURBS of order $p$ (or higher) exactly, i.e. the optimized weights $w_i$ in NURBS will be all pairwise equal (typically one). Similarly, rational functions can be fitted exactly by NURBS (only). In this example, we show that an exponential function can be fitted very well even with small number of control points and we get better fit by NURBS than B-splines. Let $f(\xi)=\operatorname{e}^{-\xi}$, $\xi\in[0,6]$, and let us consider $p=3$ and the open knot vector $C$ of length $m=13$ that is linearly spanned between $c_1 = 0$ and $c_{m} = 6$, no knots are repeated, i.e. $n = 9$. Then 
$\hat{\epsilon}_{\text{bs}} \doteq \texttt{2.7751e-05}$ and 
$\hat{\epsilon}_{\text{nrb}} \doteq \texttt{5.8463e-07}$, i.e. even for this small $n$ we can get a five decimal digits precision with B-spline and almost two degrees better accuracy by NURBS. The fitted curve and optimized weights can be seen in Figure \ref{fig:nurbs_fit_exp}.
\begin{figure}[ht!]
\includegraphics[height=60mm]{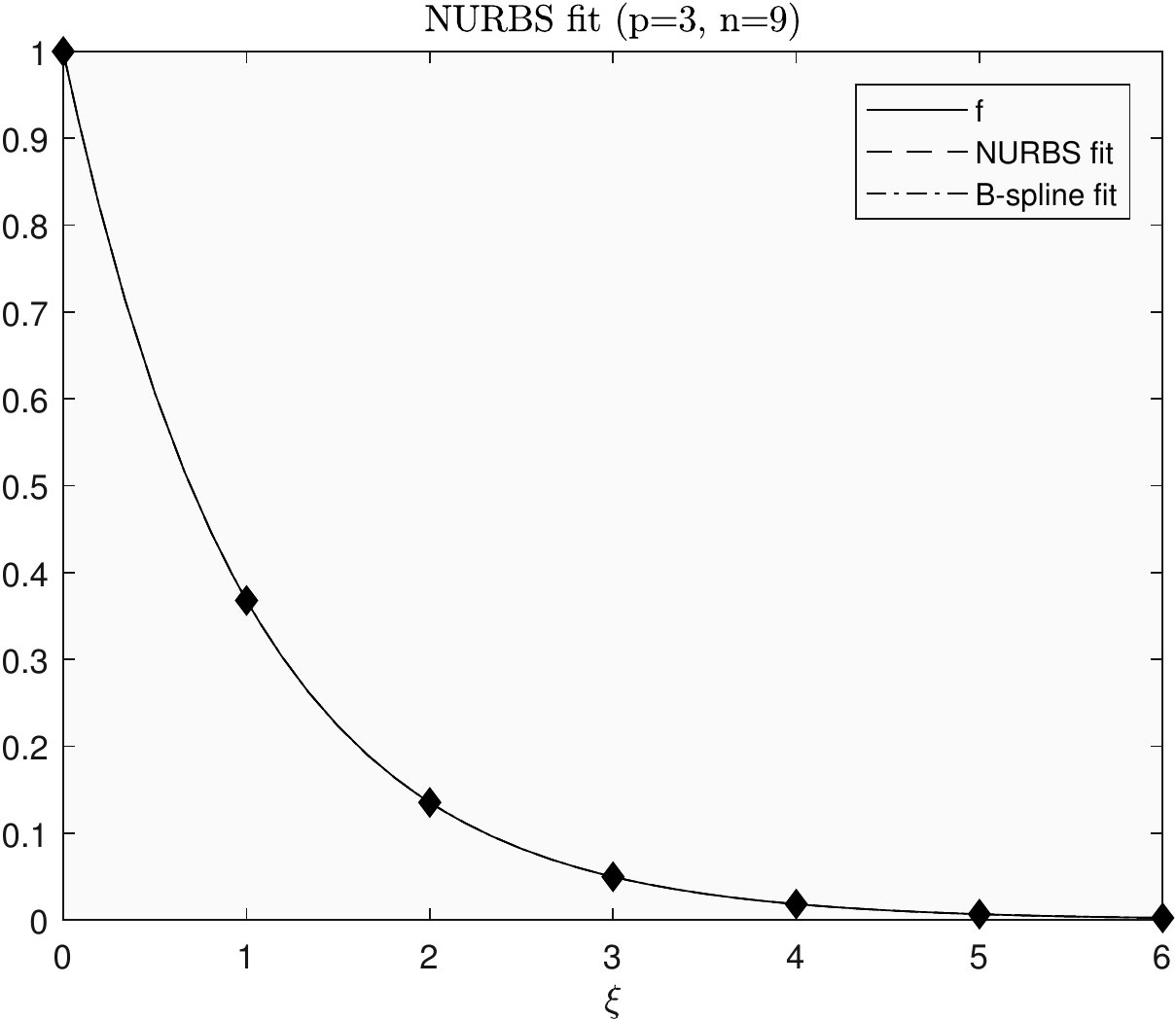}
\includegraphics[height=60mm]{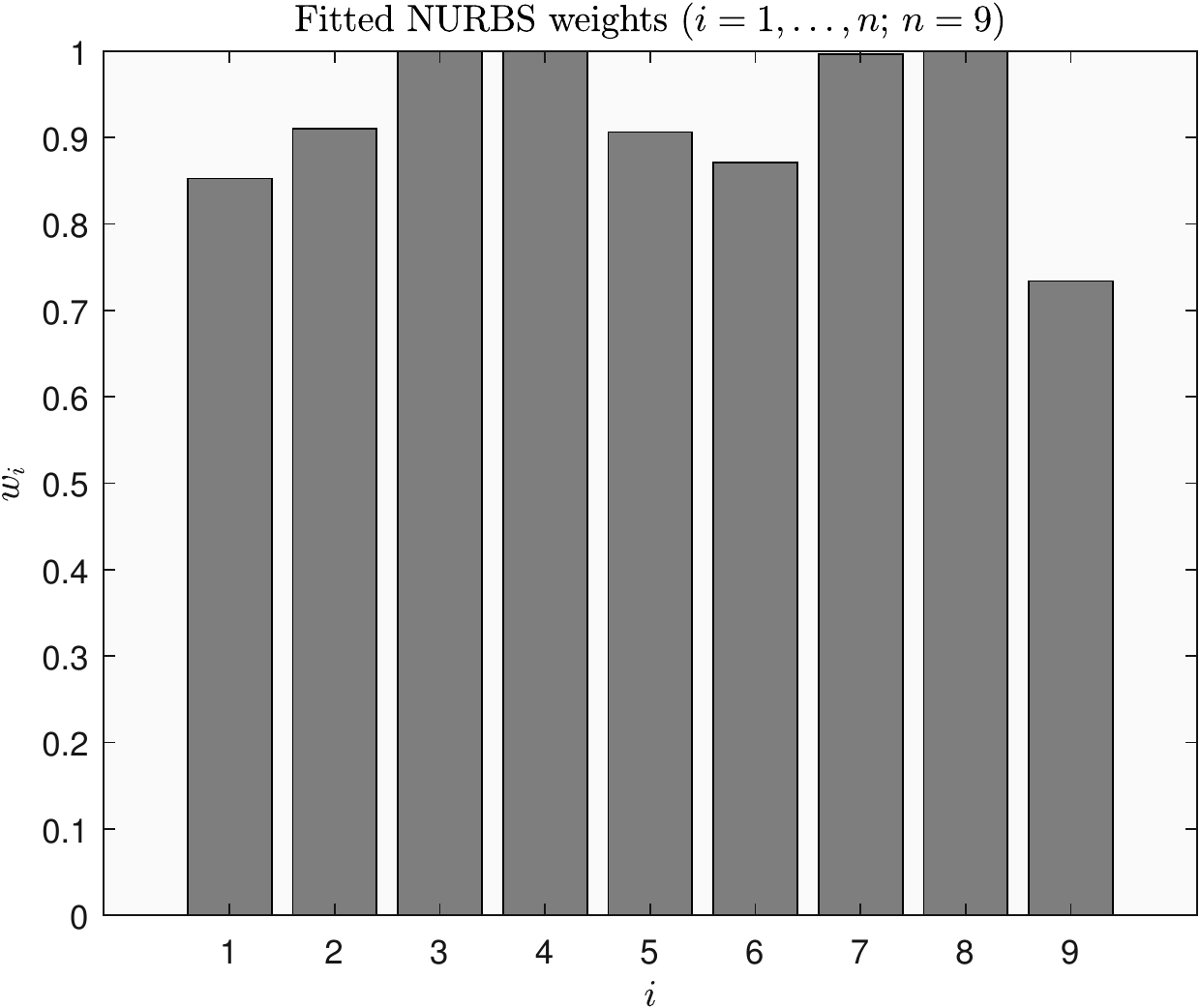}
\caption{Fitting cubic ($p=3$) B-spline and NURBS to the function $f(\xi)=\operatorname{e}^{-\xi}$, $\xi\in[0,6]$, from Example \ref{ex:nurbs_fit_exp}. In the left picture both fitted curves overlap with the original curve, the diamonds represent the corresponding knots. On the right we depict distribution of the optimized NURBS weights.}
\label{fig:nurbs_fit_exp}
\end{figure}

\end{example}

\begin{example}[NURBS fit to a non-smooth but continuous function]\label{ex:nurbs_fit_put}
In option pricing, typical payoff functions are non-smooth. Let $f(\xi) = \max(3-\xi,0)$, $\xi\in[0,6]$, that serves as an example of the pay-off function for European put option. If we use the same smooth basis from the previous example (i.e. if no knot is repeated) we are in fact trying to fit the linear combination of smooth basis functions to a non-smooth function which results in the so-called ``smoothing effect'', i.e. the fit is far from being perfect, see Figure \ref{fig:nurbs_fit_put}, we get 
$\hat{\epsilon}_{\text{bs}} \doteq \texttt{6.3688e-03}$ and 
$\hat{\epsilon}_{\text{nrb}} \doteq \texttt{2.8930e-05}$, i.e. the NURBS fit is still quite good compared to the B-spline fit. However, by repeating the knot at $\xi=3$ three times (in general $p$ times), we get the non-smooth basis which can give us a perfect fit (both parts of the graph of $f$ are in fact polynomials of degree 1). In Figure \ref{fig:nurbs_fit_put} we can see that the smooth NURBS fit is again visually indistinguishable from the original function whereas the B-spline fit is far from the original function.
\begin{figure}[ht!]
\includegraphics[height=60mm]{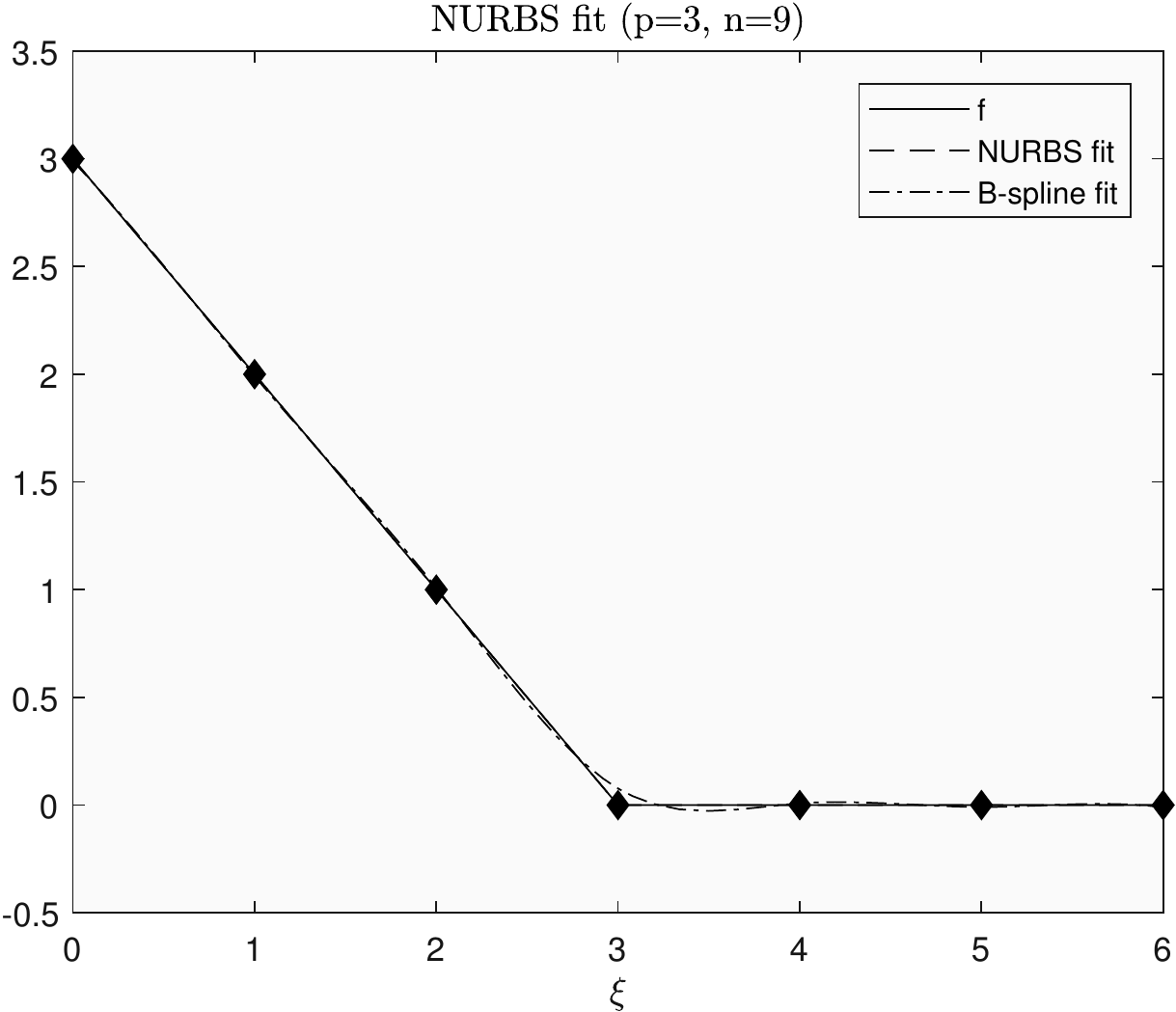}
\includegraphics[height=60mm]{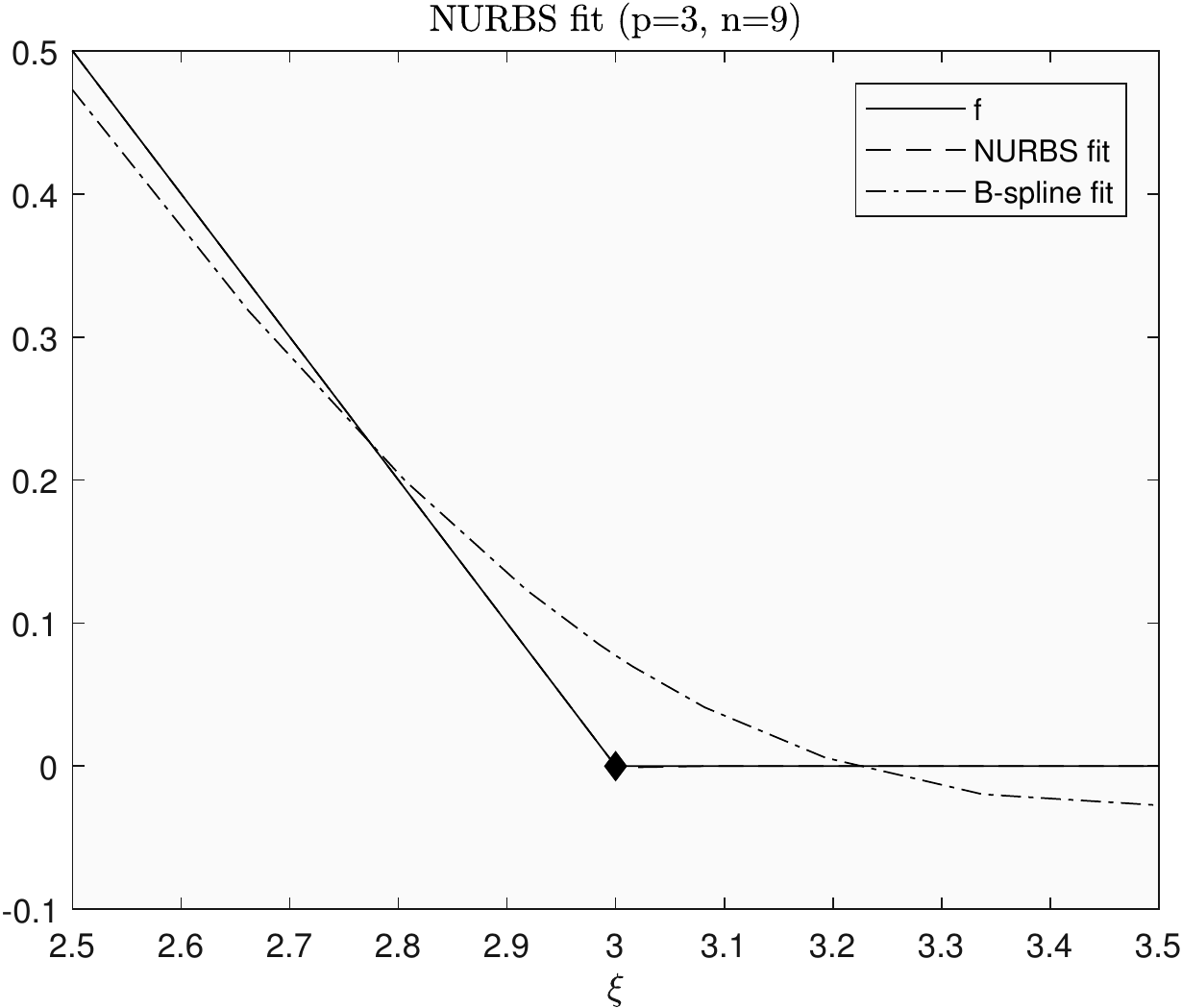}
\caption{Fitting cubic ($p=3$) B-spline and NURBS with smooth basis functions to the non-smooth function $f(\xi)=\max(3-\xi,0)$, $\xi\in[0,6]$, from Example \ref{ex:nurbs_fit_put}. To demonstrate the smoothing effect, we show the zoom of the graphs in picture on the right.}
\label{fig:nurbs_fit_put}
\end{figure}
\end{example}

\begin{example}[NURBS fit to a discontinuous function]\label{ex:nurbs_fit_digital}
Sometimes the pay-off functions can be even discontinuous, for example in digital options. Let $f(\xi) = 1$ for $\xi\geq 3$ and zero otherwise, $\xi\in[0,6]$. If we use the same smooth basis as above, the accuracy of the B-spline fit gets even worse, 
$\hat{\epsilon}_{\text{bs}} \doteq \texttt{5.6854e-02}$, however, with NURBS we still get reasonably good results 
$\hat{\epsilon}_{\text{nrb}} \doteq \texttt{4.7959e-04}$. To get the perfect fit, we can just simply repeat the knot at $\xi=3$ four times (in general $p+1$ times).

\begin{figure}[ht!]
\includegraphics[height=60mm]{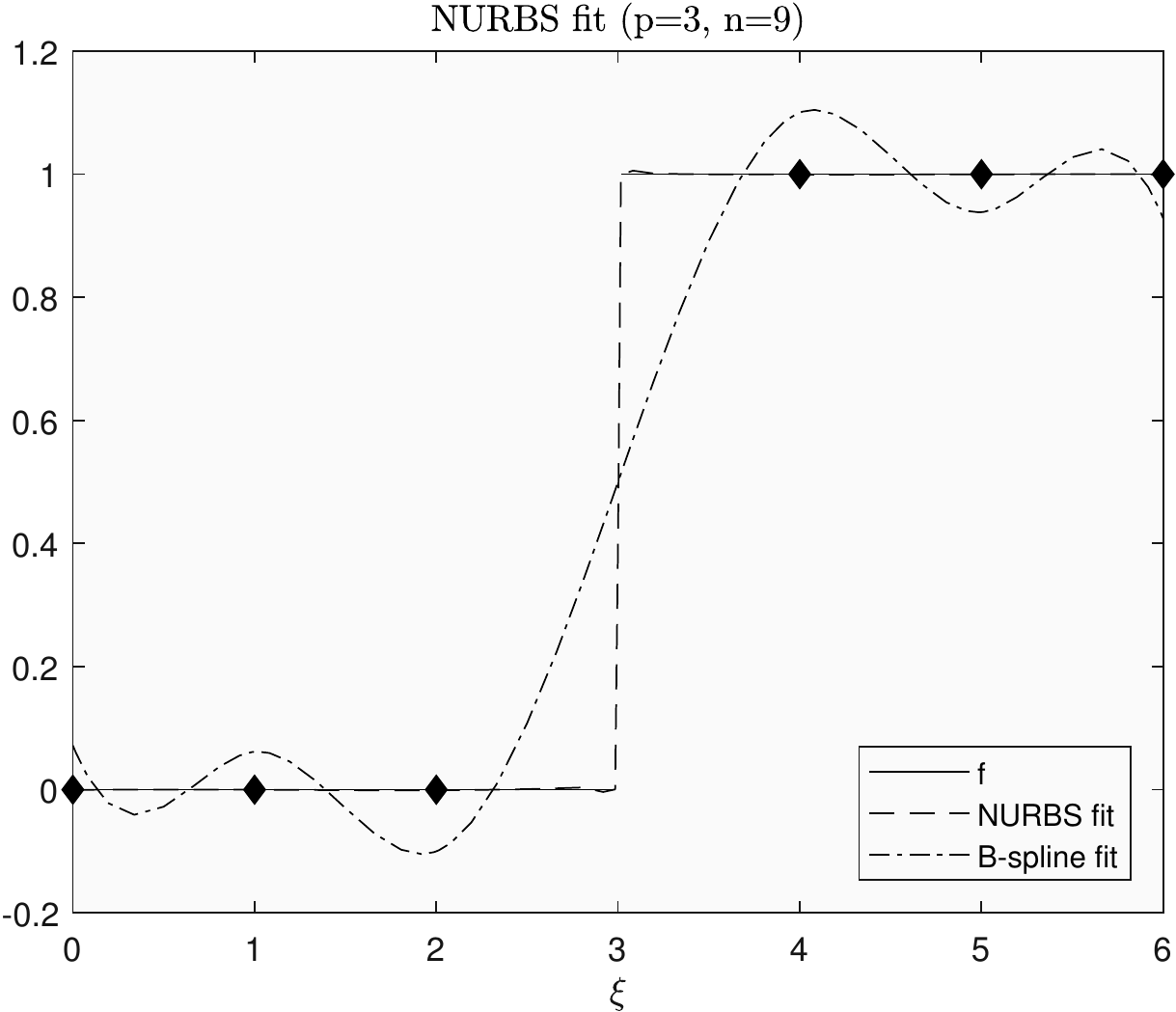}
\includegraphics[height=60mm]{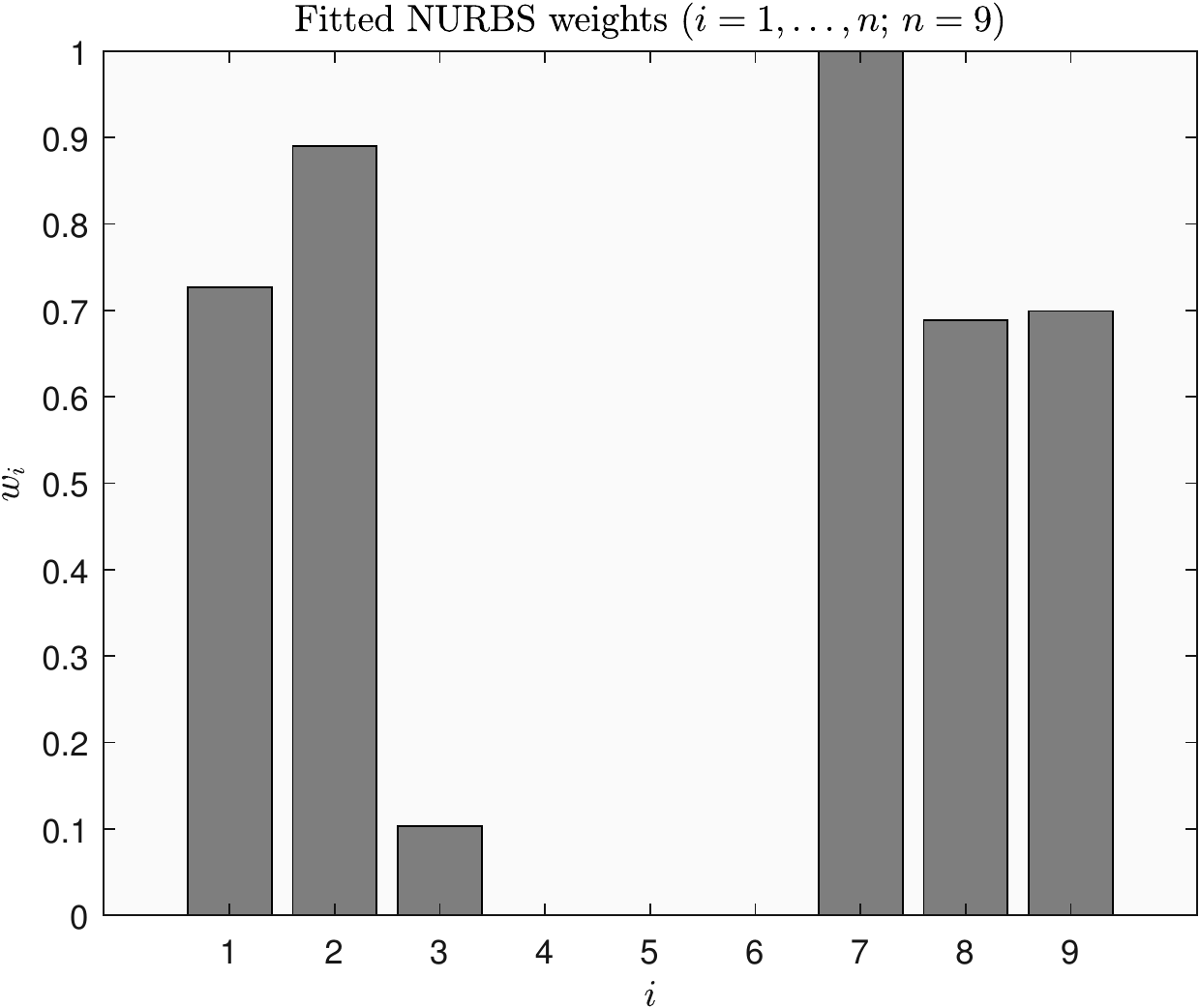}
\caption{Fitting cubic ($p=3$) B-spline and NURBS with smooth basis functions to the discontinuous function $f(\xi)$, $\xi\in[0,6]$, from Example \ref{ex:nurbs_fit_digital}. To avoid confusion, the knot at $\xi=3$ is not plotted. Fitted NURBS weights are in picture on the right. Note that weights $w_4, w_5$ and $w_6$ are very small, but not zero.}
\label{fig:nurbs_fit_digital}
\end{figure}

\end{example}

In differential equations with one spatial variable, we will, in fact, look for a solution described as the two-dimensional NURBS curve \eqref{eq:NURBS-curve} with control points $\mathbf{P}_i=[x_i,y_i], i=1,2,\dots,n$, where their $x$ coordinates are given as a partition of a finite interval $[a,b]\subset\mathbb{R}$, i.e. $a = x_1 < x_2 < \dots < x_n = b$. To solve the equation means to find the $y$ coordinates of the control points. The expression \eqref{eq:NURBS-curve} give us a transformation between the parametric coordinate $\xi$ and physical coordinates $[x,y]$. Similarly in higher dimensions. For the sake of simplicity, we do not define NURBS \emph{surfaces} or \emph{solids} here, in our problems described by evolution equations, time variable is discretized and in each time step of the iterative scheme a finite element solution (a~NURBS curve) is found.

\subsection{Option pricing models}

Following \cite{BaustianMrazekPospisilSobotka17asmb}, we consider a general SVJD model which covers several kinds of stochastic volatility processes and also different types of jumps
\begin{align*}
\mathrm{d}S_t &= (r-\lambda\beta)S_t \mathrm{d}t + \sqrt{v_t}S_t \mathrm{d}W^S_t + S_{t-} \mathrm{d}Q_t,\\
\mathrm{d}v_t &= p(v_t) \mathrm{d}t + q(v_t) \mathrm{d}W^v_t,\\
\mathrm{d}W^S_t \mathrm{d}W^v_t &= \rho\,\mathrm{d}t,
\end{align*}
where $p,q\in C^{\infty}(0,+\infty)$ are general coefficient functions,
$r$ is the risk-free interest rate, $\rho$ is the correlation of Wiener processes $W_t^S$ and $W_t^v$,
parameters $\lambda$ and $\beta$ correspond to a specific jump process $Q_t$, which is
a compound Poisson process $Q_t = \sum\limits_{i=1}^{N_t} Y_i$, where
$Y_1, Y_2, \dots$ are pairwise independent random variables with identically distributed jump sizes $\beta=\mathbb{E}[Y_i]$ for all $i\in\mathbb{N}$, $N_t$ is a standard Poisson process with intensity $\lambda$ independent of the $Y_i$.

For simplicity, we consider only the case when the jump sizes are log-normal, i.e. when $\ln(1+Y_i) \sim \mathcal{N}(\mu_J,\sigma_J^2)$ and $\beta = \exp\left\{ \mu_J + \frac12 \sigma_J^2 \right\}-1$. These types of jumps occur in original models by \cite{Merton76,Bates96,PospisilSobotka16amf}. Other types of jumps may require a simple modification, for example, for log-uniform jump sizes $\ln(1+Y_i) \sim \mathcal{U}(a,b)$, $\beta = \frac{e^b-e^a}{b-a}-1$ studied by \cite{YanHanson06}, or slightly more advanced changes for general exponential L\'{e}vy processes.

The problem of pricing an option in a model with jumps corresponds to a PIDE, see for example \cite{Hanson07}, Theorem 7.7. Let $K$ be the strike price, $\tau$ be the time to maturity and $f=f(\tau,s,v)$ denote the European option price that will be considered a function of $\tau$, price $s=S_t$ of the underlying asset (at time $t$) and volatility $v=v_t$ (at time $t$). Pricing PIDE for the European option price is of the form
\begin{align}
- f_\tau =
&-\frac{1}{2} v s^2 f_{ss} - \rho q(v) \sqrt{v} s f_sv - \frac{1}{2} q^2(v) f_{vv} \notag\\
&- (r-\lambda\beta)s f_s - p(v)f_v+r f \label{e:PIDE} \\
&- \lambda \int_0^{+\infty} \left[ f(\tau,s y,v)-f(\tau,s,v) \right] \varphi(y) \; \mathrm{d}y, \notag
\end{align}
where $f_\tau, f_s, f_{ss}, f_v, f_{vv}$ and $f_{sv}$ are corresponding partial derivatives and $\varphi(y)=\frac1{y \sigma_J\sqrt{2\pi}}\exp\left\{-\frac{(\log(y)-\mu_J)^2}{2\sigma_J^2} \right\}$ denotes the log-normal density. Initial and boundary conditions will be described in detail below. In this paper, we consider two major examples covering several models.

\begin{example}[Merton model]\label{ex:Merton}
We get a constant volatility jump diffusion model introduced by \cite{Merton76}, if we take $p(v)=q(v)=0$, i.e. if $v_t=\sigma^2$ and $\sigma$ is the constant volatility parameter. In this case it is sufficient to consider the function $f=f(\tau,s)$ as a function of one \emph{spatial} variable only and the equation \eqref{e:PIDE} reduces significantly and a numerical solution will be considered separately, see section \ref{ssec:1d} below.
Note that the \cite{BlackScholes73} model can be obtained as a special case if there are no jumps, i.e. if $Q_t=0$. 

In the top left picture of Figure \ref{fig:nurbs_fit_merton}, we can see the European put option price for Merton model with parameter values 
$r=  0.048$, 
$K = 100$, 
$T =1$, 
$\sigma = 0.197$, 
$\lambda = 0.74$, 
$\mu_J = -0.055$, 
$\sigma_J = 1.1$.
\end{example}

\begin{example}[SVJD model]\label{ex:SVJD}
We consider an approximative fractional SVJD model introduced by \cite{PospisilSobotka16amf}. In this model, volatility process is the approximative fractional Brownian motion, i.e. $p(v)=(H-1/2)\psi_t\sigma\sqrt{v}+\kappa(\theta-v)$ and $q(v)=\varepsilon^{H-1/2}\sigma\sqrt{v}$, where $H\in[1/2,1)$ is the Hurst parameter, $\varepsilon\to0$ is the approximation parameter and $\psi_t=\int_0^t (t-s+\varepsilon)^{H-3/2} \mathrm{d}W^{v}_s$. If we take $H=1/2$, we get the \cite{Bates96} model as a special case. Popular \cite{Heston93} model can be further obtained from the Bates model if there are no jumps, i.e. again if $Q_t=0$. 

SVJD price in Figure \ref{fig:nurbs_fit_svjd} is calculated for parameters
$r = 0.0529$, 
$K = 100$, 
$T =1$, 
$v_0 = 0.1$,
$\kappa = 0.5$, 
$\theta = 0.19$, 
$\sigma = 0.51$, 
$\rho = 0.24$, 
$\lambda = 0.09$, 
$\mu_J = -0.1$, 
$\sigma_J = 0.4$, 
$\epsilon = 0.3$, 
$H = 0.6$. 
\end{example}

\section{Methodology}\label{sec:methodology}

\subsection{One-dimensional problem}\label{ssec:1d}

Let us now consider the Merton model described in Example \ref{ex:Merton}. Since $v=\sigma^2$ is a constant, all partial derivatives $f_v, f_{vv}$ and $f_{sv}$ in \eqref{e:PIDE} are zero and our problem reduces to the PIDE with one spatial variable only, i.e. we are looking for a solution $f=f(\tau,s)$.

Required theory for variation methods for evolutionary problems can be found in \cite{DautrayLions92} in Chapters XVIII and corresponding numerical methods in \cite{DautrayLions93}, Chapter XX, or in more details in \cite{Trangenstein13}.

\subsubsection{Variation formulation and Discretization}

Dealing with the Merton model, we arrive at the problem of solving the PIDE
\begin{align}
\left\{
\begin{array}{l l}
f_\tau(\tau,s) - &\\
\quad - \frac{1}{2} \sigma^2 s^2 f_{ss}(\tau,s) - (r - \lambda\beta) s f_s(\tau,s) +rf(\tau,s) \,~ & \tau \in \left(0,+\infty \right)\, ,  \\
\quad - \lambda \int_0^{+\infty} \left[ f(\tau,s y)-f(\tau,s) \right] \varphi(y) \; \mathrm{d}y  =0\, ,&s \in \left(0,+\infty \right)\, , \\
f(\tau,s) = h'_D(\tau)\, ,& s\in \Gamma'_D\, , \\
f_s(\tau,s) = h'_N(\tau)\, ,& s\in \Gamma'_N\, , \\
f(0,s) = \phi(s) \, ,
\end{array}
\right.
\label{eq:met10}
\end{align}
where $\varphi$ is the log-normal density function. The set $\emptyset \neq \Gamma'_D \subset \left\{0, +\infty \right\}$ is the part of the boundary where the Dirichlet boundary condition is required. We define $\Gamma'_N \subset \left\{0, +\infty \right\}$ for Neumann boundary condition analogously. The boundary conditions at infinity are regarded as a limit behaviour of the solution. Let the boundary conditions and the initial condition be consistent. For the sake of clarity, we omit writing the dependencies of the option price $f$ where not needed. Originally, the option price is of the form $f:\left[ 0, +\infty \right) \times \left[ 0, +\infty \right) \to \mathbb{R}_0^+$. In order to be able to treat Equation \eqref{eq:met10} numerically, we restrict the spatial and the time domain and assume $s \in \left[0,\bar{s} \right], \, 0<K<\bar{s}$ and $\tau \in \left[ 0, T \right]$, $\tau>0$. Such restriction is called localization. Pricing the European put option, we arrive at the equation
\begin{align}
\left\{
\begin{array}{l l}
f_\tau(\tau,s) - &\\
\quad - \frac{1}{2} \sigma^2 s^2 f_{ss}(\tau,s) - (r - \lambda\beta) s f_s(\tau,s) +rf(\tau,s)\,~ & \tau \in \left(0,T\right) \, ,\\
\quad - \lambda \int_0^{+\infty} \left[ f(\tau,s y)-f(\tau,s) \right] \varphi(y) \; \mathrm{d}y  =0\, , & s \in \left(0,\bar{s} \right) \, , \\
f(\tau,s) = h_D(\tau)\, ,& s\in \Gamma_D \, , \\
f_s(\tau,s) = h_N(\tau)\, ,& s\in \Gamma_N \, , \\
f(0,s) = \phi(s) \, ,
\end{array}
\right.
\label{eq:met15}
\end{align}
where $\emptyset \neq \Gamma_D \subset \left\{0, \bar{s} \right\}$ and $\Gamma_N \subset \left\{0, \bar{s} \right\}$. The first equation in \eqref{eq:met15} can be expressed in a form
\begin{align}
f_\tau - \left( P f_s \right)_s + Q f_s + R f + J(f) = 0,
\end{align}
where 
\begin{align}
P &= \frac{1}{2} \sigma^2 s^2 \, ,\\
Q &= (\lambda\beta - r + \sigma^2)s \, ,\\
R &= r \, ,\\
J(f) &= - \lambda \int_0^{+\infty} \left[ f(\tau,s y)-f(\tau,s) \right] \varphi(y) \;  \mathrm{d}y \, .
\end{align}
The problem of finding solution of \eqref{eq:met15} is reformulated to finding a function $f(\tau, \, \cdot \,) \in W_D^{1,2}(0,\bar{s})$ for all $\tau \in \left[0,T\right]$ \footnote{The space $W_D^{1,2}(0,\bar{s})$ is the space of functions satisfying 
\begin{itemize}
\item $f(\tau,s) = h_D(\tau), \quad s\in \Gamma_D$ (Dirichlet boundary condition),
\item $\left\Vert f(t) \right\Vert_{L^2([0,\bar{s}])}^2 + \left\Vert f_s(t) \right\Vert_{L^2([0,\bar{s}])}^2  < +\infty $ where the derivative is considered in the weak sense. 
\end{itemize}} and $f_\tau(\, \cdot \, , s)$ existing in the classical sense for all $s \in \left[0,\bar{s}\right]$ such that 
\begin{multline}
\int_0^{\bar{s}} f_\tau g \;  \mathrm{d}s + \int_0^{\bar{s}} \left( f_s P g_s + f_s Q g + f R g \right) \; \mathrm{d}s\, - \\
- \left[ f_s P g \right]_{s \in \Gamma_N} + \int_0^{\bar{s}} J(f) g \; \mathrm{d}s = 0 \, , 
\label{eq:met20}
\end{multline}
holds for all $g \in W_D^{1,2}([0,\bar{s}])$ and $\tau \in [0, T]$. The term $\left[ f_s P g \right]_{s \in \Gamma_N}$ means sum oriented with respect to the direction of the outer normal of the interval $[0,\bar{s}]$, i.e. the term at 0 is subtracted provided $0 \in \Gamma_N$ and the term at $\bar{s}$ is added provided $\bar{s} \in \Gamma_N$. 

Let us define a set of basis functions $\Psi_N = \left( \psi_1, \psi_2, \ldots, \psi_N \right)^T$ such that $\psi_i \in W^{1,2}([0,\bar{s}])$, $\sum_{i=1}^N \psi_i^2(s) \neq 0$ for all $s\in \left[ 0, \bar{s} \right]$ and $\psi_i$ are pairwise linearly independent. Let us assume that the there exists exactly one basis function $\psi_i$ for each $s \in \Gamma_D$ such that the function is nonzero at the Dirichlet point. Let us denote $I = (1,2, \ldots, N)^T$ the set of all indices, $I_D$ the set of indices of the basis functions corresponding to the Dirichlet points and $\overline{I_D}$ the complement of $I_D$ with respect to $I$. In what follows, we will derive all the formulations using a general basis functions $\psi_i$, but the core idea of the isogeometric analysis is to take $\psi_i$ to be the NURBS basis function $R_i^p$ defined in \eqref{eq:NURBS-basis}. Since the NURBS basis functions are defined in the parametric space and we are looking for the solution in the real (physical) space, in all calculations below we must keep in mind that there is always the transform between these two, i.e. the integrals below already contain the corresponding \emph{Jacobian}. This is one of the biggest differences to the classical FEM where basis functions are already considered in the real space variables. 

The principle of the Galerkin FEM is to solve~\eqref{eq:met20} at each time step in the space $H_N$ defined $H_N := \sum_{i\in I_D} f_i \psi_i + \mathrm{span}\{ \Psi_{N,\overline{I_D}} \}$. Note that $f_i$ for $i \in I_D$ are uniquely determined. The space $H_N$ is a finite-dimensional subspace of the Hilbert space $W_D^{1,2}([0,\bar{s}])$ and thus it is sufficient to consider \eqref{eq:met20} valid for $g \in \Psi_N$. Let us assume $f^N(\tau), f_\tau^N(\tau) \in H_N$ for a given time $\tau \in [0,T]$ such that
\begin{align}
f^N(\tau,s) &:= \mathbf{f}^T(\tau) \cdot \Psi_N(s) = \sum_{i=1}^N f_i(\tau) \psi_i(s) \in H_N \, \\
f_\tau^N &:= \mathbf{f}_\tau^T(\tau) \cdot \Psi_N(s) = \sum_{i=1}^N f_{i,\tau}(\tau) \psi_i(s) \in H_N.
\end{align}
Then, the finite-dimensional approximation of~\eqref{eq:met20} in the space $H_N$ can be expressed as a finite system of ODEs
\begin{align}
\left\{
\begin{array}{rcl}
\mathbb{M} \cdot \mathbf{f}_\tau(\tau) + \mathbb{A} \cdot \mathbf{f}(\tau) + \mathbb{J} \cdot \mathbf{f}(\tau) &=& \mathbf{b} \, , \\
\mathbb{M} \cdot \mathbf{f}(0) &=& \mathbf{\Phi} \, , \\
\end{array}
\right.
\label{eq:met30}
\end{align}
with the initial condition 
\begin{align}
\Phi = \left( \int_0^{\bar{s}} \psi_1 \phi \; \mathrm{d}s, \int_0^{\bar{s}} \psi_2 \phi \; \mathrm{d}s, \ldots, \int_0^{\bar{s}} \psi_N \phi \; \mathrm{d}s \right)^T \, ,
\end{align} 
that is approximated in the $L^2$ sense in the space $H_N$ and where
\begin{align}
\mathbb{M} &= \left( \int_0^{\bar{s}} \psi_j \psi_i \; \mathrm{d}s \right)_{i,j=1,\ldots,N}\, ,\\
\mathbb{A} &= \Biggl(\int_0^{\bar{s}} \Biggl( \frac{1}{2} \sigma^2  s^2 \left( \psi_j \right)_s \left( \psi_i \right)_s 
+ (\lambda\beta - r + \sigma^2) s \psi_i (\psi_j)_s + r \psi_j \psi_i \Biggr) \mathrm{d}s \Biggr)_{i,j}\, , \label{eq:met33} \\
\mathbb{J} &= \left( \int_0^{\bar{s}} \left( -\lambda \int_0^{+\infty} \left( \left( \psi_j(sy)-\psi_j(s) \right) \psi_i(s) \varphi(y) \right) \mathrm{d}y \right) \mathrm{d}s \right)_{i,j} \, , \label{eq:met34}\\
\mathbf{b} &= \left( \left[ ( h_N P \psi_{j} \right]_{s \in \Gamma_N} \right)_{j}^T \, .
\label{eq:met35}
\end{align}
The matrices $\mathbb{M}$ and $\mathbb{A}$ are called mass and stiffness matrices respectively. With a proper choice of the basis $\Psi_N$ we say that the problem~\eqref{eq:met30} is in the semidiscretized form. We discretize the system~\eqref{eq:met30} along the time axis equidistantly with the time step $d\tau$, i.e. let $\tau_i = i\,d\tau, i=0,1,\dots,n_\tau$ and substitute the time derivative by a backward difference 
\begin{align}
f_\tau(\tau_i) \approx \frac{f(\tau_i)-f(\tau_{i-1})}{d\tau}, 
\end{align}
to introduce a fully discrete weighted scheme of the first order. We obtain
\begin{align}
\left\{
\begin{array}{rl}
\multicolumn{1}{l}{\mathbb{M} \cdot \frac{\mathbf{f}(\tau_{i})-\mathbf{f}(\tau_{i-1})}{d\tau} + \omega (\mathbb{A}+\mathbb{J}) \cdot \mathbf{f}(\tau_{i}) + (1-\omega) (\mathbb{A} + \mathbb{J}) \cdot \mathbf{f}(\tau_{i-1})+} &= \mathbf{b} \, , \\
\mathbb{M} \cdot \mathbf{f}(\tau_0) &= \mathbf{\Phi} \, ,
\end{array}
\right.
\label{eq:met40}
\end{align}
where $i=1,\ldots,n_\tau$ and $\omega \in \left[ 0, 1 \right]$ is the weight of the time scheme. Note that the jump term is weighted. This can make computational expenses higher for the sake of better accuracy. For an explicit treatment of the jump term see e.g.~\cite{Feng07}. The system of equations~\eqref{eq:met40} can be rewritten as
\begin{align}
\left\{
\begin{array}{rcl}
\left(\mathbb{M} + {d\tau}  \omega  (\mathbb{A}+\mathbb{J}) \right) \cdot \mathbf{f}(\tau_{i}) &=&\left( \mathbb{M} - (1-\omega)  d\tau   (\mathbb{A}+\mathbb{J}) \right) \cdot \mathbf{f}(\tau_{i-1}) + {d\tau} \mathbf{b} \, , \\
\mathbb{M} \cdot \mathbf{f}(\tau_0) &=& \mathbf{\Phi} \, .
\end{array}
\right.
\label{eq:met50}
\end{align}
If $\omega=0$ or $\omega=1$, we say that the scheme~\eqref{eq:met50} is fully explicit or implicit, respectively.

\subsubsection{Implementation -- European put option}
When pricing the European put option, we consider the boundary conditions
\begin{align}
h_D(\tau) &= K e^{-r \tau} \, , &\quad \Gamma_D(s) = \{0\} \, ,  \\
h_N(\tau) &= 0 \, , &\quad \Gamma_N(s) = \{\bar{s} \} \, , 
\end{align}
and the initial condition
\begin{align}
\phi(s) &=  \max \{K-s,0 \} =: (K-s)^+ \, .
\end{align}
Thus, the scheme~\eqref{eq:met50} is simplified, since $\mathbf{b} = 0$.
 
In order to handle the solution of the problem~\eqref{eq:met15} numerically, we discretize the spatial and the time domain. Thus, let $\mathbf{s} = (s_0,s_1, \ldots, s_{n_s})^T$ such that $0 = s_0 < s_1 < \ldots < s_{n_s} = \bar{s}$ and $\mathbf{\tau} = (\tau_0, \tau_1, \ldots, \tau_{n_\tau})^T$ such that $0 = \tau_0 < \tau_1 < \ldots < \tau_{n_\tau}= T$. In the further text, we assume an equally spaced discretization of the time domain with the step $d\tau$. Note that B-spline and NURBS basis satisfies the linearity independence and boundary properties posed in the previous paragraph and therefore the described discretization procedure can be used. The matrices $\mathbb{A}$ and $\mathbb{M}$ can be computed in a straightforward manner. Computing the jump matrix $\mathbb{J}$ is done by substituting $y = x/s$ in the double integral term in \eqref{eq:met34}
\begin{align}
\int_{0}^{\bar{s}} & \left( -\lambda \int_0^{+\infty} \left( \left( \psi_j(sy)-\psi_j(s) \right) \psi_i(s) \varphi(y) \right) \mathrm{d}y \right) \mathrm{d}s = \\
 = &\lambda \int_0^{\bar{s}} \psi_j(s) \psi_i(s) \; \mathrm{d}s \underbrace{\int_0^{+\infty} \varphi(y) \; \mathrm{d}y}_{=1} -  \nonumber\\
 &- \lambda \int_0^{\bar{s}} \left( \int_0^{+\infty} \psi_j(sy)  \psi_i(s) \varphi(y)  \; \mathrm{d}y \right) \mathrm{d}s\ , \\
 =& \lambda \int_0^{\bar{s}} \psi_j(s) \psi_i(s) \; \mathrm{ds} - \nonumber \\
 &- \lambda \int_0^{\bar{s}} \int_0^{\bar{s}} \psi_{j}(x) \psi_{i}(s) \varphi \left( \frac{x}{s} \right) \frac{1}{s} \; \mathrm{d}x \, \mathrm{d}s \, .
\end{align}
The Fubini's theorem was implicitly used since all the functions and their products are bounded and integrable in their respective integration regions. The second integral term in the second line is equal to one since $\varphi$ is assumed to be a probability density function defined on $[0, +\infty)$. The matrix $\mathbb{J}$ can be expressed in the form
\begin{align}
\mathbb{J} = \lambda (\mathbb{M} -\mathbb{J}') \, ,
\label{eq:met55}
\end{align}
where 
\begin{align}
\mathbb{J}' = \left( \int_0^{\bar{s}} \int_0^{\bar{s}} \psi_{j}(x) \psi_{i}(s) \varphi \left( \frac{x}{s} \right) \frac{1}{s} \;  \mathrm{d}x \, \mathrm{d}s \right)_{i,j = 1,2, \ldots, N}\, .
\end{align}
Naturally, the domain truncation leads to an error in computing the integral term. The value of the integral term over the interval $[\bar{s},+\infty)$ can be estimated for the European call option, for further details see~\cite{Almendral05}. Note that the asymptotical estimate of the European put price for high underlying values is close to zero.  

Finally, we incorporate the Dirichlet boundary condition by a direct assignment. The function $\psi_1$ is the only basis function nonzero at the boundary $s=0$ and the equality $\psi_1(0)=1$ holds. Thus, the corresponding value of $f_1$ is equal directly to the value of the boundary condition at each time step
\begin{align}
f_1(\tau_i)= K e^{-r \tau_i},
\label{eq:met60}
\end{align}
for $i = 0,1, \ldots, n_\tau$. For the sake of simplicity, let us denote $I = (1,2,\ldots, N)^T$ and $\overline{I_D} = (2,3,\ldots, N)^T$ the indices not corresponding to the term $f_1$. The initial condition of the iterative scheme~\eqref{eq:met50} remains intact but the recursion equation collects the known terms on the right handside of the equality. Thus, we have
\begin{multline}
\left(\mathbb{M}_{\overline{I_D},\overline{I_D}} + d\tau \omega (\mathbb{A}_{\overline{I_D},\overline{I_D}}+\mathbb{J}_{\overline{I_D},\overline{I_D}}) \right) \cdot \mathbf{f}_{\overline{I_D}}(\tau_{i}) = \\ 
= \left( \mathbb{M}_{\overline{I_D},I} - (1-\omega) d\tau  (\mathbb{A}_{\overline{I_D},I}+\mathbb{J}_{\overline{I_D},I})\right) \cdot \mathbf{f}_I(\tau_{i-1}) - \\
-(\mathbb{M}_{\overline{I_D},1}+ d\tau \; \omega (\mathbb{A}_{\overline{I_D},1}+\mathbb{J}_{\overline{I_D},1})) f_1(\tau_i) \, , 
\end{multline}
where the subscripts $I, \overline{I_D}$ denote corresponding submatrices and subvectors. The price of the European call option is later computed via put-call parity, i.e.
\begin{align}
f_C(\tau,s) = f_P(\tau,s) + s - K e^{-r \tau} \, ,
\end{align}
where $f_C$ (or $f_P$) is the price of the European call (or put) option at time $t = T -\tau$ with the price of underlying equal to $s$. 

\begin{example}\label{ex:met1}
Let us consider the Merton model with parameters from Example \ref{ex:Merton}
and with parameters for the numerical solution $n_\tau =100$, $n_s = 9$, $\bar{s} = 300$. 
In Figure~\ref{fig:met1}, we can see the European put price calculated using the FEM with the B-spline basis of degree 3. The comparison of the numerical solution to the closed-form solution is depicted in the picture on the right. The influence of the truncation error can be easily seen at $s$ values close to $\bar{s}$. The choice of the Neumann boundary condition at $\bar{s}$ is suitable for obtaining an accurate result for a relatively small value of $n_s$ and $\bar{s}$. However, the convergence of such scheme for $\bar{s} \to +\infty$ is for this type of boundary condition unclear. 

\begin{figure}[htbp]
\begin{tabular}{p{0.5\textwidth}p{0.5\textwidth}}
\vspace{0pt}\includegraphics[height=60mm,trim={4mm 0mm 0mm 4mm},clip]{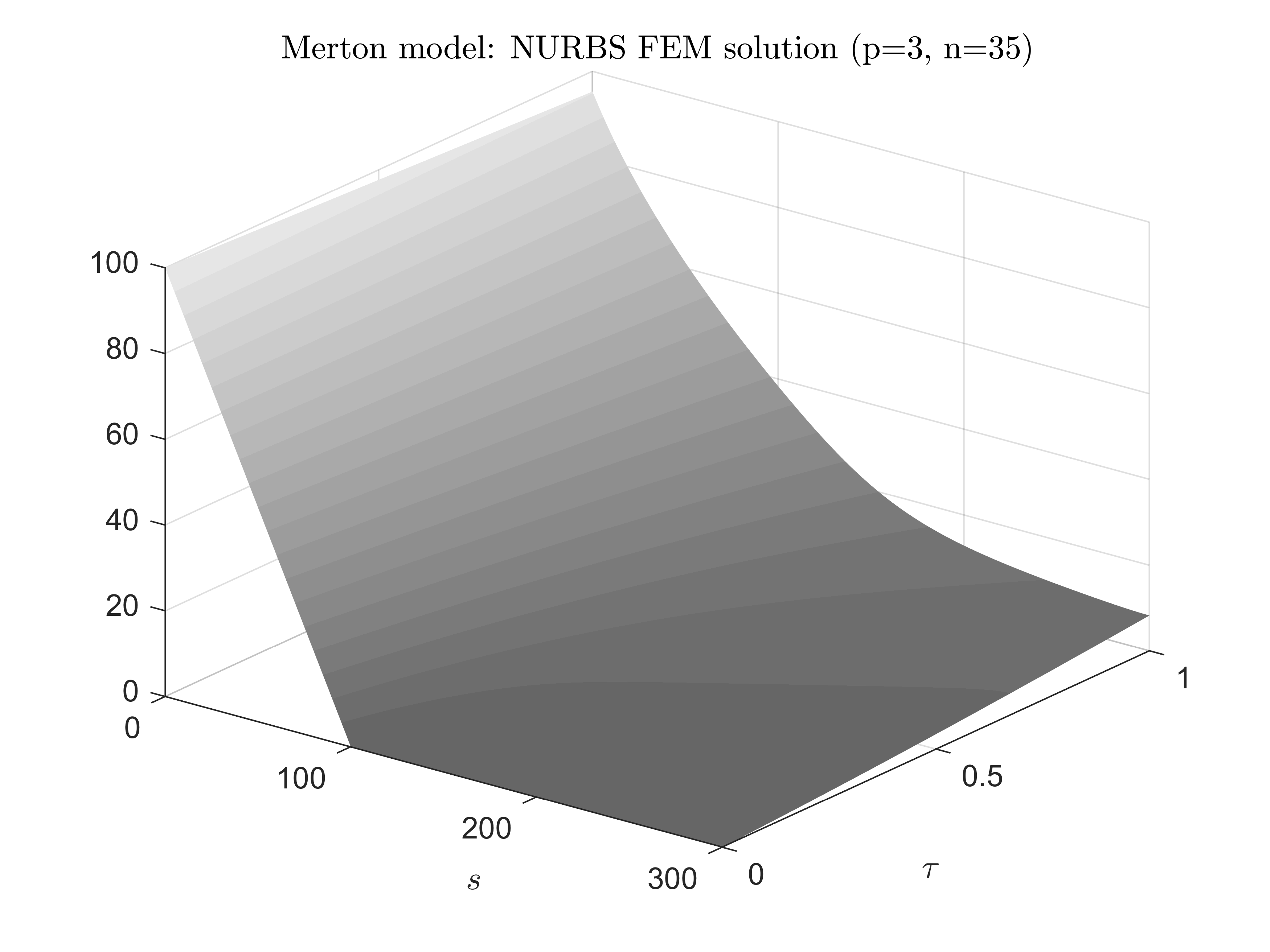}&
\vspace{0pt}\includegraphics[height=58mm]{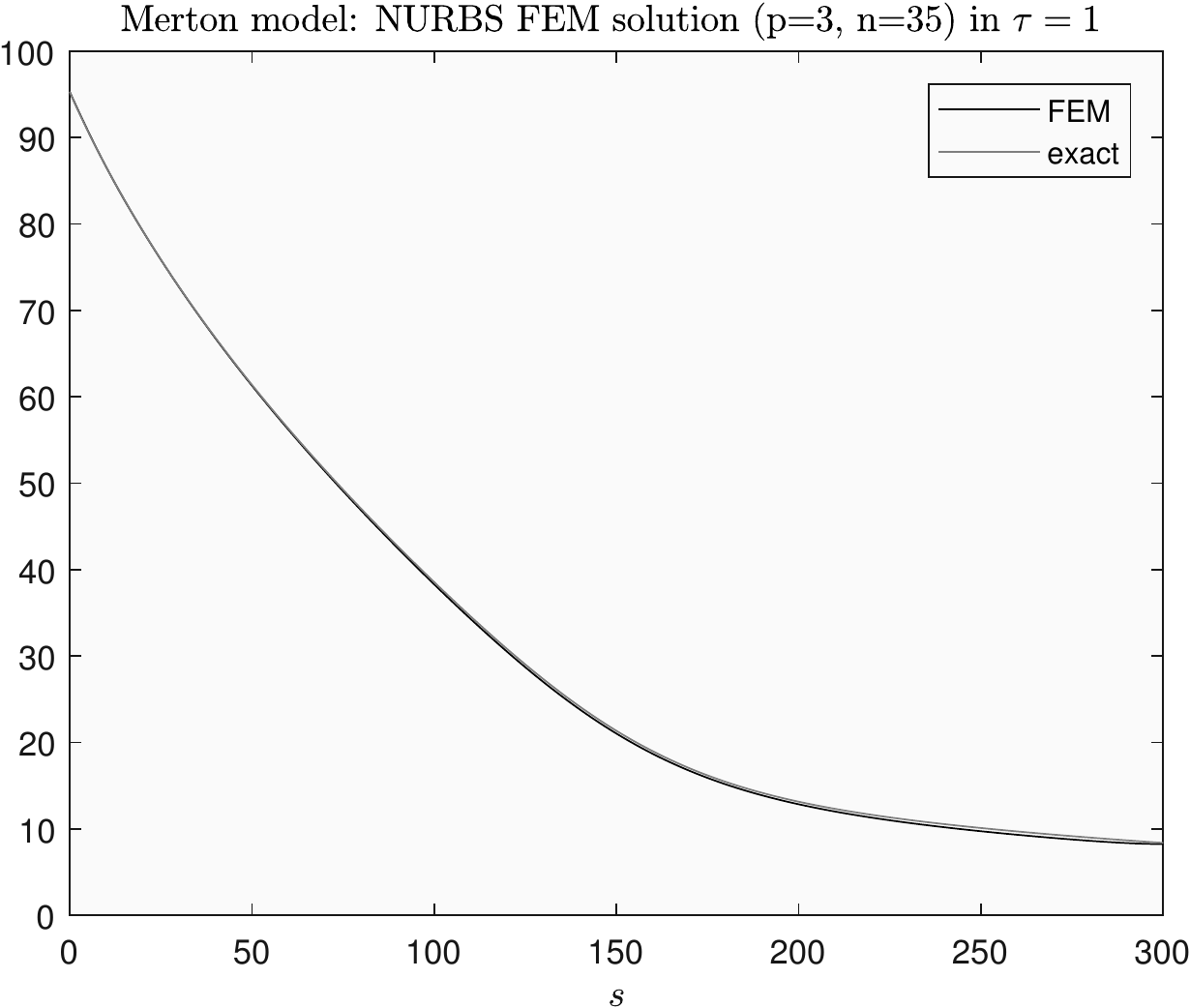}
\end{tabular}
\caption{Example of pricing European put option by the Merton model with parameters listed in Example~\ref{ex:met1}.}
\label{fig:met1}
\end{figure}
\end{example}

\subsection{Multidimensional problem}\label{ssec:2d}
\subsubsection{Variation Formulation and Discretization}
Price of the European option in the SVJD model from Example \ref{ex:SVJD} satisfies the PIDE
\begin{align}
\left\{
\begin{array}{l l}
f_\tau -                                                                                  & \tau \in \left(0, +\infty \right)\, ,\\
\quad -\frac{1}{2} v s^2 f_{ss} - \rho q(v) \sqrt{v} s f_sv - \frac{1}{2} q^2(v) f_{vv}  & s \in \left(0, +\infty \right)\, ,\\
\quad  - (r - \lambda\beta)s f_s - p(v)f_v+r f -                                          & v \in \left(0, +\infty \right)\, , \\
\quad - \lambda \int_0^{+\infty} \left[ f(\tau,s y,v)-f(\tau,s,v) \right] \varphi(y) \; \mathrm{d}y  =0 \,~ , & \\
f(\tau,s,v) = h'_D(\tau,s,v)\, ,& (s,v) \in \Gamma'_D \, , \\
\nabla f(\tau,s,v)\, \cdot \overrightarrow{n_P} = h'_N(\tau,s,v) \, ,& (s,v)\in \Gamma'_N \, , \\
f(0,s,v) = \phi(s,v) \, .
\end{array}
\right.
\label{eq:met65}
\end{align}
Where $\overrightarrow{n_P} := P \, \cdot \, \overrightarrow{n}$ is the matrix product of matrix $P$ (see~\eqref{eq:met73}) and $\overrightarrow{n}$ is the outer normal of the region $\left(0, +\infty \right) \times \left(0, +\infty \right)$. Once again, the localization $f: [0,T] \times [0,\bar{s}] \times [0,\bar{v}] \to \mathbb{R}_0^+$ is performed and thus,~\eqref{eq:met65} is reformulated
\begin{align}
\left\{
\begin{array}{l l}
f_\tau - \,  \\
\quad -\frac{1}{2} v s^2 f_{ss} - \rho q(v) \sqrt{v} s f_{sv} - \frac{1}{2} q^2(v) f_{vv} & \tau\in \left(0, T \right)\, ,\\
\quad  - (r- \lambda\beta)s f_s - p(v)f_v+r f - & s \in (0,\bar{s}) \, , \\
\quad - \lambda \int_0^{+\infty} \left[ f(\tau,s y,v)-f(\tau,s,v) \right] \varphi(y) \; \mathrm{d}y  =0 \,~ , & v \in (0,\bar{v}) \, ,\\
f(\tau,s,v) = h_D(\tau,s,v)\, ,& (s,v)\in \Gamma_D \, , \\
\nabla f(\tau,s,v)\, \cdot \overrightarrow{n_P} = h_N(\tau,s,v)\, ,& (s,v)\in \Gamma_N \, , \\
f(0,s,v) = \phi(s,v) \, .
\end{array}
\right.
\label{eq:met70}
\end{align}
The first expression in~\eqref{eq:met70} can be written in the form
\begin{align}
f_\tau - \nabla \cdot P \nabla f + Q^T \cdot \nabla f + R f + J(f)= 0, 
\end{align}
where
\begin{align}
P &= \frac{1}{2}
\left(
\begin{array}{cc}
v s^2 & \rho q(v) \sqrt{v} s \label{eq:met73}\\
\rho q(v) \sqrt{v} s & q^2(v)
\end{array}
\right) \, , \\
Q &= \left( 
\begin{array}{c}
-(r - \lambda\beta)s+ v s +\frac{1}{2} \rho s q'(v) \sqrt{v} + \frac{1}{4} \rho s \frac{q(v)}{\sqrt{v}} \\
-p(v)+\frac{1}{2} \rho q(v) \sqrt{v} + q(v)q'(v) 
\end{array}
\right) \, , \\
R &= r \, , \\
J(f) &= - \lambda \int_0^{+\infty} \left[ f(\tau,s y,v)-f(\tau,s,v) \right] \varphi(y) \;  \mathrm{d}y \, .
\label{eq:met75}
\end{align}
Let us denote $\Omega := (0,\bar{s}) \times (0,\bar{v})$ and $\mathbf{x} := (s,v)^T$. The variation formulation of the problem~\eqref{eq:met70} is searching for a function  $f \in W_D^{1,2}(\Omega)$ \footnote{Sobolev space of functions satisfying Dirichlet boundary condition in~\eqref{eq:met70} in the sense of traces.} such that 
\begin{multline}
\int_\Omega f_\tau g \; \mathrm{d\mathbf{x}} + \int_\Omega \left( \nabla f \cdot P \cdot \nabla g + (Q^T \cdot \nabla f)g + R  f  g \right) \mathrm{d\mathbf{x}} - \\
- \int_{\Gamma_N} \left( \nabla f \cdot P \cdot \overrightarrow{\mathbf{n}} \right) g \; \mathrm{d\mathbf{S}} + \int_\Omega J(f)  g \; \mathrm{d\mathbf{x}} = 0 \, ,
\label{eq:met80}
\end{multline}
holds for all $g \in W_D^{1,2}(\Omega)$ and all $\tau \in [0, T]$. As in the one-dimensional case, we define a set of basis functions $\Psi_N = \left( \psi_1, \psi_2 \ldots, \psi_N \right)^T$ such that $\psi_i \in W^{1,2}(\Omega)$, $\sum_{i=1}^N \psi_i^2(s) \neq 0$ for all $\mathbf{x} \in \Omega$ and $\psi_i$ are pairwise linearly independent. Further, we assume that the restriction of the basis functions nonzero on the Dirichlet boundary $\Gamma_D$ to the boundary $\Gamma_D$ is pairwise linearly independent. Indices of such basis functions will be denoted by $I_D$, set of indices of all basis functions will be denoted $I$ and complement of $I_D$ in $I$ will be denoted $\overline{I_D}$.  We can find an unique $\mathbf{f}_{I_D} \in \mathbb{R}^{|I_D|}$ such that the function $\mathbf{f}_I^T \cdot \Psi_{N.I_D} + \mathbf{a}^T \cdot \Psi_{N,\overline{I_D}}$ satisfies given Dirichlet boundary condition for all $\mathbf{a} \in \mathbb{R}^{N-|I_D|}$ in the sense of the best $L^2$-approximation. Therefore, we search for the solution in the space $H_N := \sum_{i \in I_D} f_i \psi_i + \mathrm{span} \left\{ \Psi_{N,\overline{I_D}} \right\}$. Using the basis of $H_N$ as test functions we arrive at~\eqref{eq:met30} where
\begin{align}
\mathbb{M} =& \left( \int_\Omega \psi_j \psi_i \; \mathrm{d}\mathbf{x} \right)_{i,j=1,\ldots,N}\, ,\\
\mathbb{A} =& \left( \int_\Omega \left( \frac{1}{2} s^2 v (\psi_j)_s (\psi_i)_s +\frac{1}{2} \rho q(v) \sqrt{v}  s  (\psi_j)_s (\psi_i)_v + \right. \right. \\
& + \frac{1}{2} \rho q(v) \sqrt{v}  s   (\psi_j)_v (\psi_i)_s + \frac{1}{2} q^2(v) (\psi_j)_v (\psi_i)_v + \nonumber  \\
& + \left(-(r - \lambda\beta)s+ v s +\frac{1}{2} \rho s q'(v) \sqrt{v} + \frac{1}{4} \rho s \frac{q(v)}{\sqrt{v}} \right) (\psi_j)_s \psi_i + \nonumber \\
& \left. \left. + \left( -p(v)+\frac{1}{2} \rho q(v) \sqrt{v} + q(v)q'(v)  \right) (\psi_j)_v \psi_i + r \psi_j \psi_i \right)\mathrm{d}\mathbf{x} \right)_{i,j}\, , \nonumber\\ 
\mathbb{J} =& \left( \int_\Omega \left( -\lambda \int_0^{+\infty} \left( \left( \psi_j(sy,v)-\psi_j(s,v) \right) \psi_i(s,v) \varphi(y) \right) \mathrm{d}y \right) \mathrm{d}\mathbf{x} \right)_{i,j} \, ,\label{eq:met85} \\
\mathbf{b} =& \left( \int_{\partial \Gamma_N} (h_N \psi_{j}) \mathrm{d}\mathbf{S} \right)_{j}^T \, .
\label{eq:met90}
\end{align}
Subsequently, we arrive at iterative scheme~\eqref{eq:met50}. 

\subsubsection{Implementation -- European put option}
Let $\mathbf{s} = (s_0,s_1, \ldots, s_{n_s})^T$ such that $0 = s_0 < s_1 < \ldots < s_{n_s} = \bar{s}$ and $\mathbf{v} = (v_0, v_1, \ldots, v_{n_v})^T$ such that $0 = v_0 < v_1 < \ldots < v_{n_v}= \bar{v}$. We define two bases 
\begin{align}
\Psi^s &:= (\psi_1^s, \ldots, \psi_{N_1}^s)^T \, , \\
\Psi^v &:= ( \psi_1^v, \dots, \psi_{N_2}^v)^T \, ,
\end{align}
consisting of arbitrary spline basis in the variable $s$ and $v$, respectively which define 2-D basis consisting of functions $\psi_{i+N_1(j-1)}(s,v) := \psi_i^s(s) \psi_j^v(v)$ for $i = 1,2, \ldots, N_1$ and $j = 1,2, \ldots, N_2$. Let $\Psi^N$ be defined 
\begin{align}
\Psi_N := (\psi_1, \psi_2, \ldots, \psi_{N_1 N_2})^T
 \end{align} 
as the basis consisting of $N := N_1 N_2$ functions. Note that $\Psi^s$ and $\Psi^v$ can be bases of arbitrary degree. Integral of the product of the basis functions $\psi_i(s,v) = \psi_{i_1}^s(s) \psi_{i_2}^v(v)$ and $\psi_j(s,v) = \psi_{j_1}^s(s) \psi_{j_2}^v(v)$ can be evaluated as a product of two integrals on line segments
\begin{align}
\int_\Omega \psi_i(s,v) \psi_j(s,v) \; \mathrm{d}\mathbf{x} &= \int_\Omega \psi_{i_1}^s(s) \psi_{i_2}^v(v) \psi_{j_1}^s(s) \psi_{j_2}^v(v) \; \mathrm{d}\mathbf{x} \, , \nonumber \\
&= \int_0^{\bar{s}} \psi_{i_1}^s \psi_{j_1}^s \; \mathrm{d}s \int_0^{\bar{v}} \psi_{i_2}^v \psi_{j_2}^v \; \mathrm{d}v \, ,
\end{align}
leading to much faster construction of matrices $\mathbb{M}, \mathbb{A}$ compared to case of general meshing of $\Omega$. The entries of the matrix $\mathbb{J}$ are computed through the same idea because
\begin{align}
\int_\Omega & \left( -\lambda \int_0^{+\infty} \left( \left( \psi_j(sy,v)-\psi_j(s,v) \right) \psi_i(s,v) \varphi(y) \right) \mathrm{d}y \right) \mathrm{d}\mathbf{x} = \\
& = \int_\Omega \left( -\lambda \int_0^{+\infty} \left( \left( \psi_{j_1}^s(sy)\psi_{j_2}^v-\psi_{j_1}^s(s) \psi_{j_2}^v \right) \psi_{i_1}^s \psi_{i_2}^v \varphi(y) \right) \mathrm{d}y \right) \mathrm{d}\mathbf{x} \, , \\
& = \lambda \int_0^{\bar{v}} \psi_{j_2}^v \psi_{i_2}^v \; \mathrm{d}v \; \int_0^{\bar{s}} \psi_{j_1}^s(s) \psi_{i_1}^s(s) \; \mathrm{ds} - \nonumber \\
 & \quad - \lambda \int_0^{\bar{v}} \psi_{j_2}^v \psi_{i_2}^v \; \mathrm{d}v \; \int_0^{\bar{s}} \int_0^{\bar{s}} \psi_{j_1}^s(x) \psi_{i_1}^s(s) \varphi \left( \frac{x}{s} \right) \frac{1}{s} \; \mathrm{d}x \, \mathrm{d}s \, .
\end{align}
Denoting 
\begin{align}
\mathbb{J'} = \left( \int_0^{\bar{v}} \psi_{j_2}^v \psi_{i_2}^v \; \mathrm{d}v \; \int_0^{\bar{s}} \int_0^{\bar{s}} \psi_{j_1}^s(x) \psi_{i_1}^s(s) \varphi \left( \frac{x}{s} \right) \frac{1}{s} \; \mathrm{d}x \, \mathrm{d}s \right)_{i,j = 1, \ldots, N}
\end{align}
the matrix $\mathbb{J}$ can be expressed as in~\eqref{eq:met55}.
 
The boundary conditions in multidimensional case for pricing the European put option were chosen to be (for a slightly different approach to the boundary conditions see e.g. \cite{Hout16}) 
\begin{align}
h_D(\tau,s,v) &= K e^{-r \tau} \, , \qquad              s = 0 \, , \\ 
h_N(\tau,s,v) &= \left\{ 
\begin{array}{l l}
0   \,,            & \qquad v = 0 \, , \\
0 \, ,  & \qquad v = \bar{v} \, , \\
0 \, , & \qquad s = \bar{s} \, . 
\end{array}
\right.
\end{align}

The initial condition is a standard put pay-off function
\begin{align}
\phi(s, v) &= (K-s)^+ \, .
\end{align}
Note, that the scheme~\eqref{eq:met50} is simplified, since the only Neumann boundary condition considered is a homogeneous one ($\mathbf{b} = 0$) . Using the notation from previous section we denote $I_D$ the set of all indices of the basis functions nonzero at the boundary. At a given time $\tau$, the coefficients $\mathbf{f}_{I_D}(\tau)$ of the Dirichlet basis functions can be computed by solving the system 
\begin{align}
\mathbb{M}_D \mathbf{f}_{I_D}(\tau) = \mathbf{c}(\tau)
\label{eq:met100}
\end{align}
with 
\begin{align}
\mathbf{c} &= \left( \int_{\Gamma_D} \psi_j \vert _{\Gamma_D} h_D(\tau) \; \mathrm{d}\mathbf{S} \right)_i
\end{align} 
and $\mathbb{M}_D$ being the mass matrix of the restriction of functions $\Psi_{I_D}$ to the boundary $\Gamma_D$. Note that the system~\eqref{eq:met100} must be computed at each time step of the iterative scheme. The initial condition is approximated via solving the equation
\begin{align}
\mathbb{M}_{\mathrm{IC}} \mathbf{f}(0) = \Phi, 
\end{align}
where $\mathbb{M}_{\mathrm{IC}}=\mathbb{M}$ is the mass matrix and the vector $\Phi$ is defined as
\begin{align}
\Phi &= \left(  \int_\Omega \psi_j \phi \; \mathrm{d}\mathbf{x} \right)_{j=1,\ldots,N} \, \\
&= \left( \int_0^{\bar{v}} \psi_{j_2} \; \mathrm{d}v \int_0^{\bar{s}} \psi_{j_1} \phi \; \mathrm{d}\mathbf{s} \right)_j \, ,
\end{align}
since the initial condition is constant in the variable $v$. 

\begin{example}\label{ex:met2}
Let us consider the SVJD model with parameters from Example \ref{ex:SVJD} 
and with parameters for the numerical solution $n_s = n_v = 30$, $n_\tau =100$, $\bar{s} = 300$, $\bar{v} = 3$, $p(v) = (H- 1/2)\psi_t \sigma \sqrt{v}+ \kappa(\theta -v)$, $q(v) = \epsilon^{H-1/2}\sigma \sqrt{v}$, i.e. the volatility process is driven by an approximative fractional Brownian motion. In Figure~\ref{fig:met2}, we can see the European call price calculated using the FEM with the B-spline basis of degree 3. The surface clearly shows that the Dirichlet boundary conditions (except at $s=0$) are not suitable for the pricing on the truncated domain. The picture on the right-handside shows comparison of the FEM solution of the model and the semi-closed formula. Note, that the semi-closed formula is known only for a specific choice of the functions $p, q$. However, the framework developed above can handle the model with general $p, q$. 
\begin{figure}[htbp]
\begin{tabular}{p{0.5\textwidth}p{0.5\textwidth}}
\vspace{0pt}\includegraphics[height=60mm,trim={4mm 0mm 0mm 4mm},clip]{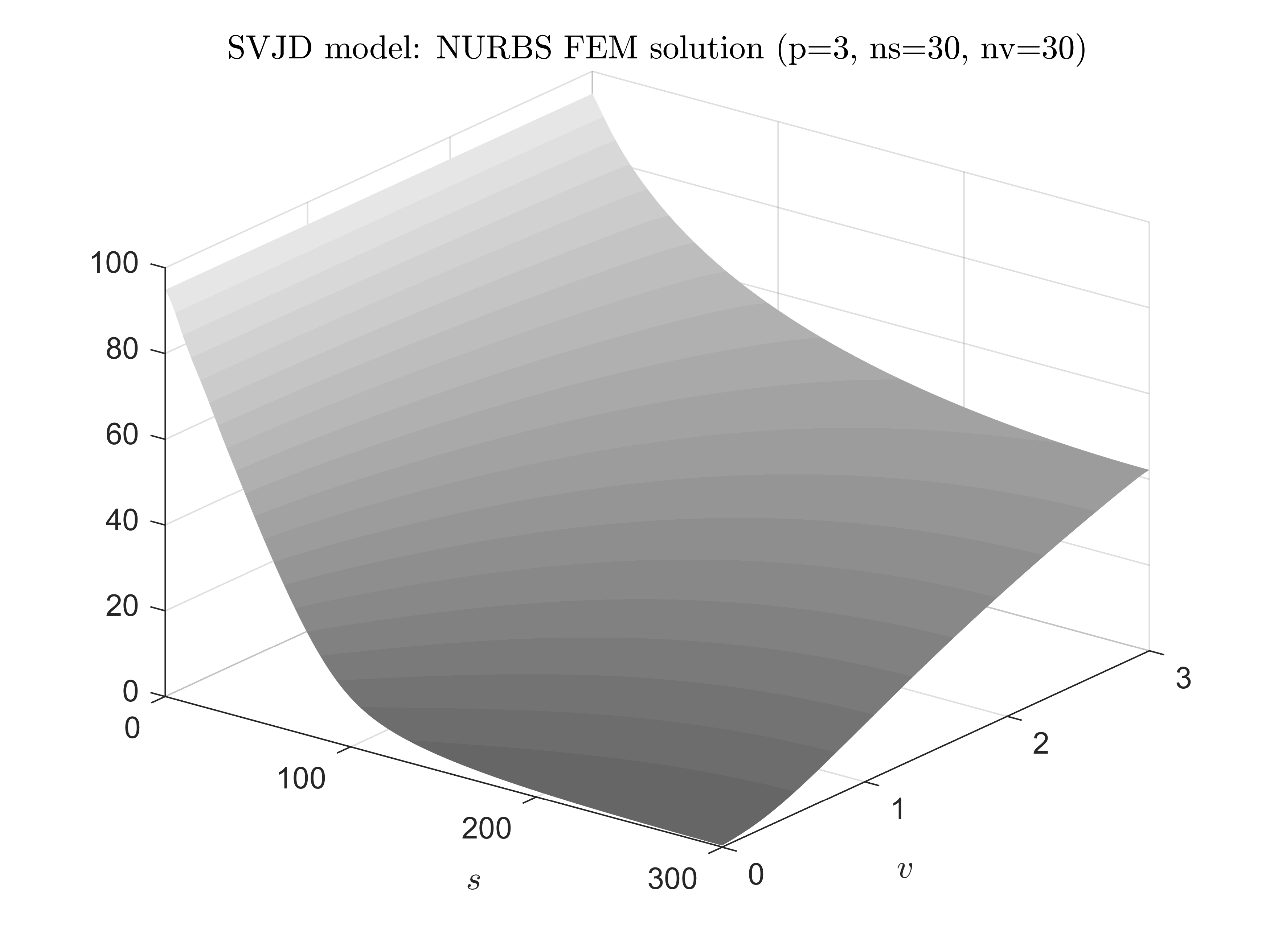}&
\vspace{0pt}\includegraphics[height=58mm]{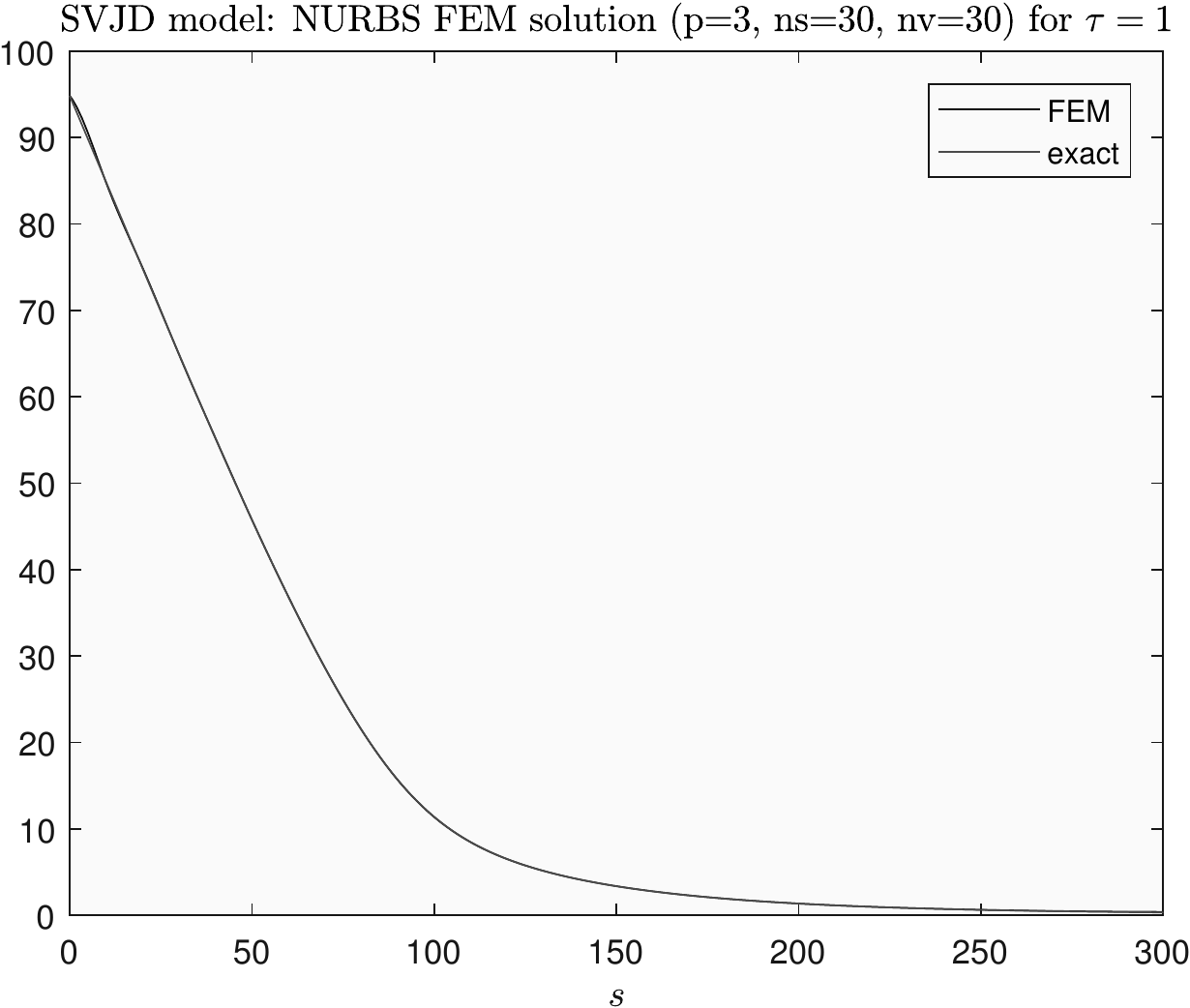}
\end{tabular}
\caption{Example of pricing European call option by the SVJD model with parameters listed in Example~\ref{ex:met2}.  }
\label{fig:met2}
\end{figure}
\end{example}

\section{Numerical results}\label{sec:results}

\subsection{Fitting exact pricing formulas by NURBS}\label{ssec:nurbs_fit}

\begin{figure}[h!]
\includegraphics[width=\textwidth]{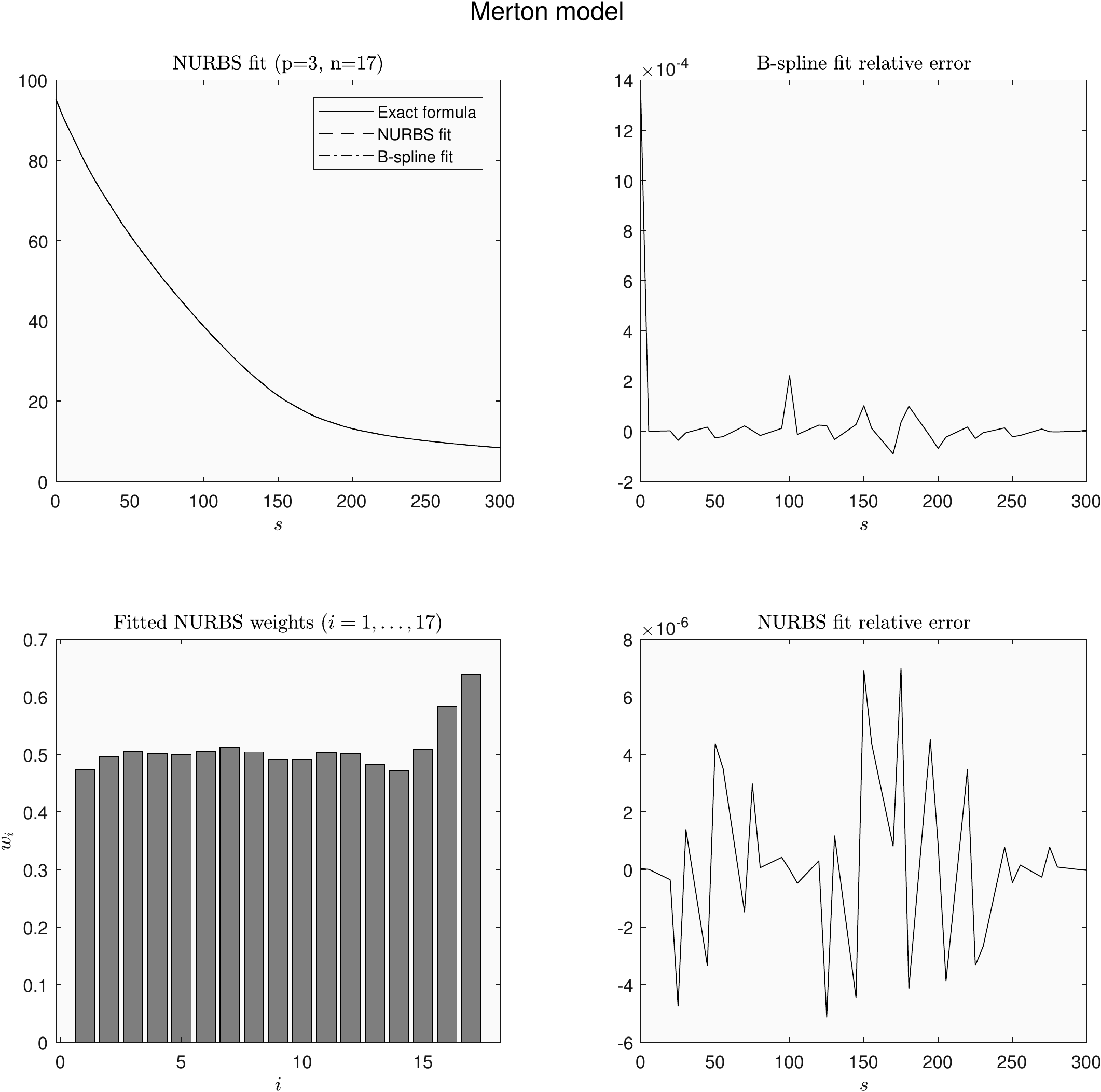}
\caption{Fitting cubic ($p=3$) B-spline and NURBS to the Merton model pricing formula, considered parameter values are from Example \ref{ex:Merton}.}
\label{fig:nurbs_fit_merton}
\end{figure}

\begin{figure}[h!]
\includegraphics[width=\textwidth]{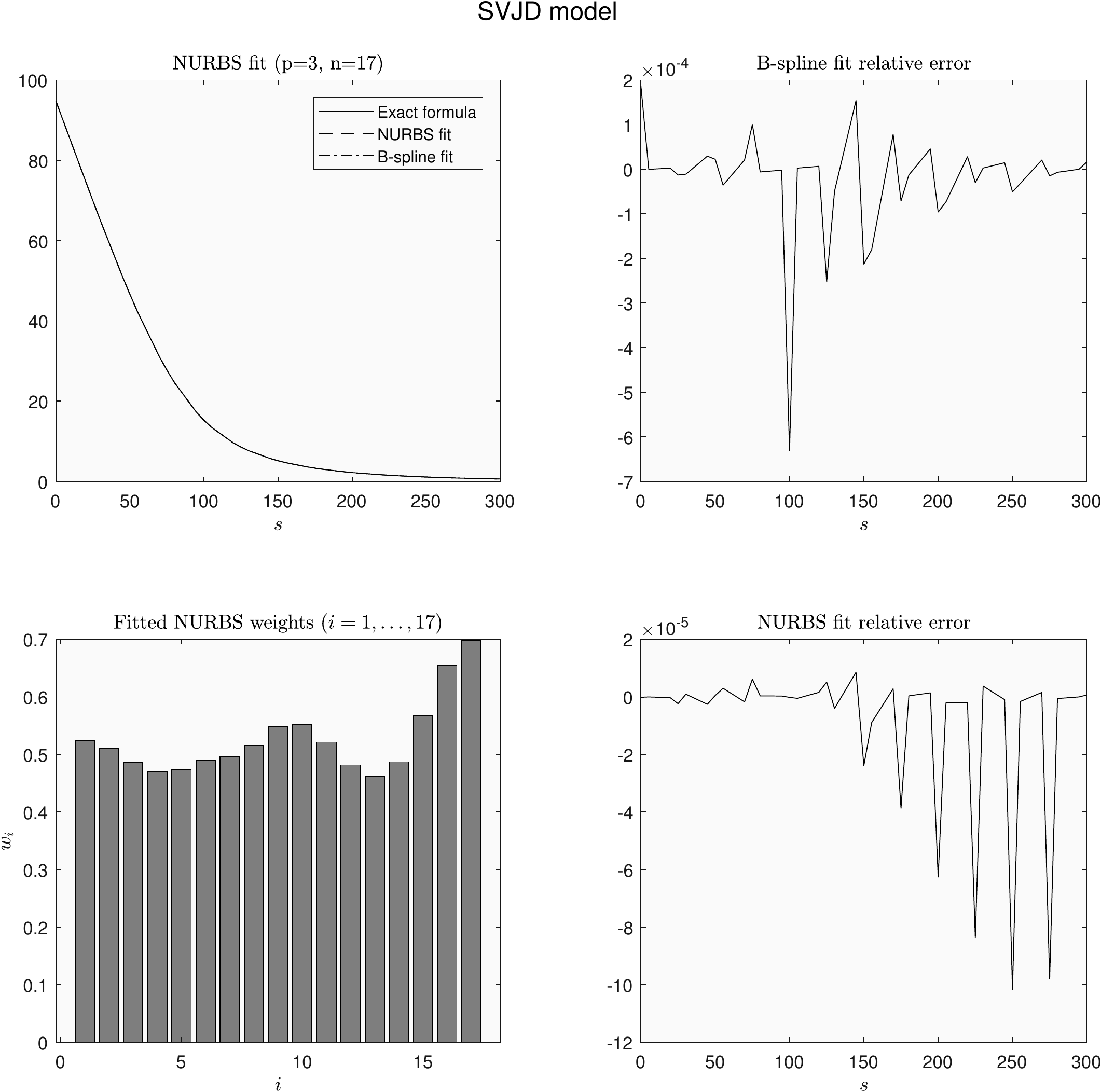}
\caption{Fitting cubic ($p=3$) B-spline and NURBS to the Merton model pricing formula, considered parameter values are from Example \ref{ex:SVJD}.}
\label{fig:nurbs_fit_svjd}
\end{figure}

For both Merton and SVJD models, there exists a semi-closed pricing formula \citep{BaustianMrazekPospisilSobotka17asmb} for the European call/put options that will be used for comparison to the numerical solution of the PIDE below. In the view of the examples of the Section \ref{sec:preliminaries}, we can fit B-spline and NURBS to this exact semi-closed pricing formula that we consider as a function $f(s)$ of the stock price $s$ if the time to maturity $\tau = T$, i.e. in time $t=0$. In Figures \ref{fig:nurbs_fit_merton} and \ref{fig:nurbs_fit_svjd} we can see the fit results for the cubic B-spline and NURBS fit to this pricing formula for the Merton and SVJD models respectively. Top left picture shows the exact formula with both fits that are visually indistiguishable. The difference is in the mean $L^2$ error that is for Merton model 
$\hat{\epsilon}_{\text{bs}} \doteq \texttt{4.0837e-05}$, 
$\hat{\epsilon}_{\text{nrb}} \doteq \texttt{4.3813e-06}$, and for the SVJD model 
$\hat{\epsilon}_{\text{bs}} \doteq \texttt{3.1733e-05}$, 
$\hat{\epsilon}_{\text{nrb}} \doteq \texttt{2.5288e-06}$. In pictures on the right we plot the relative fit error, i.e. the values $(f(s)-\hat{f}(s))/f(s)$. Bottom left picture then shows the fitted NURBS weights. For both models we use the same open knot vector, where the knot corresponding to the ATM value $s=K$ is repeated 3 times, $m=21$ and hence $n=17$. The knots therefore divide the physical coordinates interval $[0,3K]$ to $n_s = 12$ ``finite elements'' (here intervals) and the above-mentioned mean $L^2$ errors were obtained when considering only three spatial steps per element (i.e. by considering 2 Gaussian quadrature abscissas per element only).  

\begin{figure}[h!]
\includegraphics[width=0.49\textwidth]{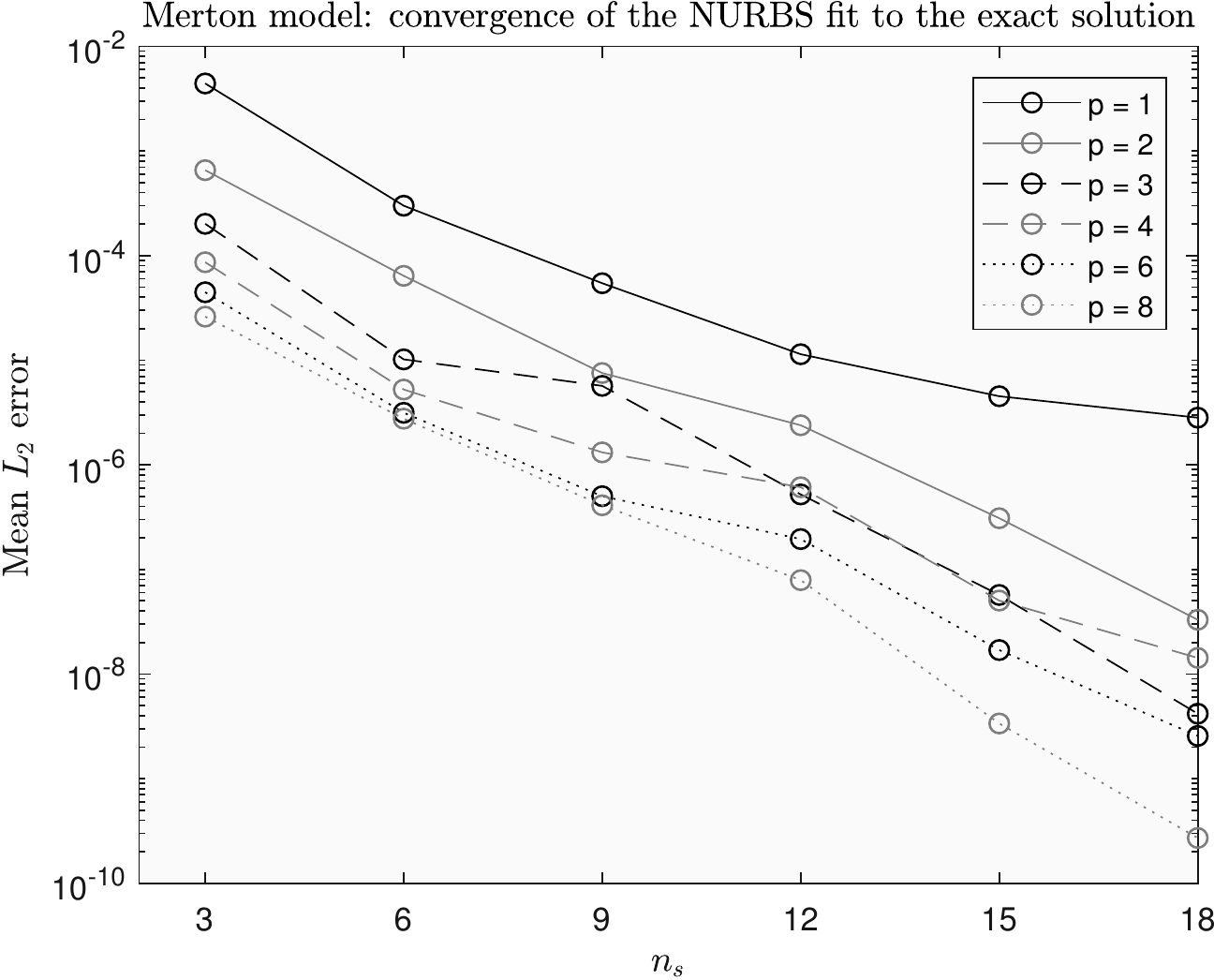}
\includegraphics[width=0.49\textwidth]{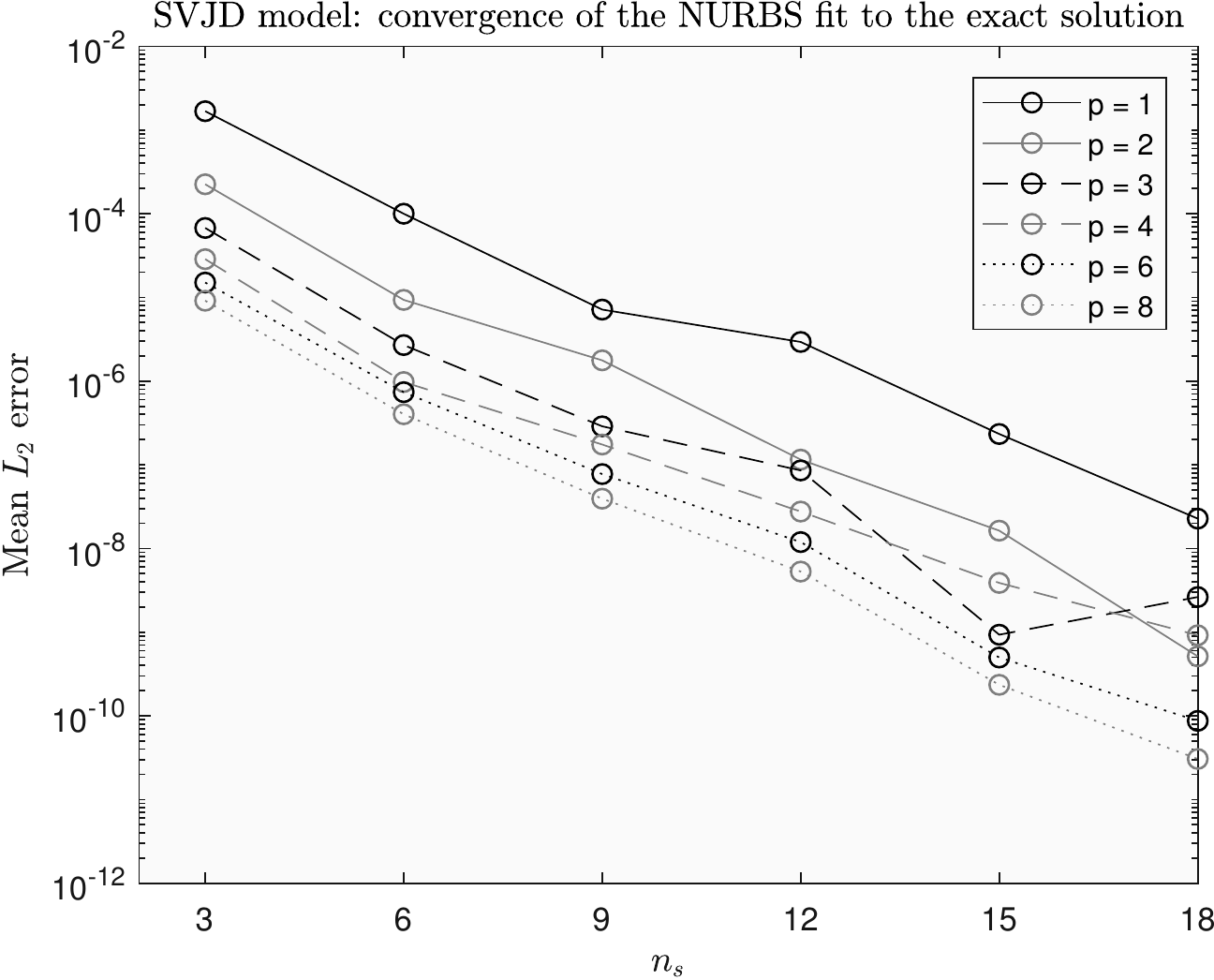}
\caption{Convergence of the NURBS fit both for the Merton and SVJD models. Mean $L^2$ error is depicted for different $p$ as a function of $n_s$.}
\label{fig:nurbs_fit_comparison}
\end{figure}

Let us now show how the mean $L^2$ error depends on the order $p$ and the number of elements $n_s$. Figure \ref{fig:nurbs_fit_comparison} confirms that for practical purposes it is sufficient to consider cubic ($p=3$) NURBS even for very small number of elements $n_s$. High accuracy for small number of discretization points is probably the biggest advantage over other fitting techniques. Using NURBS basis functions in FEM gives, therefore, a big advantage to other bases or, for example, finite differences method.

\subsection{FEM solution}\label{ssec:nurbs_fem}

We supply two examples showing the accuracy of the FEM approximations in the sense of the mean $L^2$ error. As stated above, we focus on the untransformed form of the equations~\eqref{eq:met10},~\eqref{eq:met65}. The other possibility is to introduce the substitution $X = \log(s)$ or $X = \log(\dfrac{s}{K})$ which guarantees the independence of the linear coefficients on the variable $s$. The approach presented in this manuscript is accompanied by its advantages and disadvantages. First, we are able to describe initial conditions (pay-off functions) exactly (note that the nonzero part of the transformed call and put payoff is an exponential function (see Example~\ref{ex:nurbs_fit_exp}). On the other hand, tackling the numerical analysis and obtaining strong numerical results is much more complicated due to the spatial dependence of the linear coefficients. 

In both examples, we use a homogeneous open knot vector with knot at strike multiplied $p$-times which leads to the non-smoothness of the basis at $K$ (see Figure~\ref{fig:nurbs-basis}) in the $s$ variable. The basis is smooth in the $v$ variable in the second example.  

\begin{figure}[ht!]
\centering
\begin{subfigure}{.48\textwidth}
\includegraphics[width=.99\linewidth]{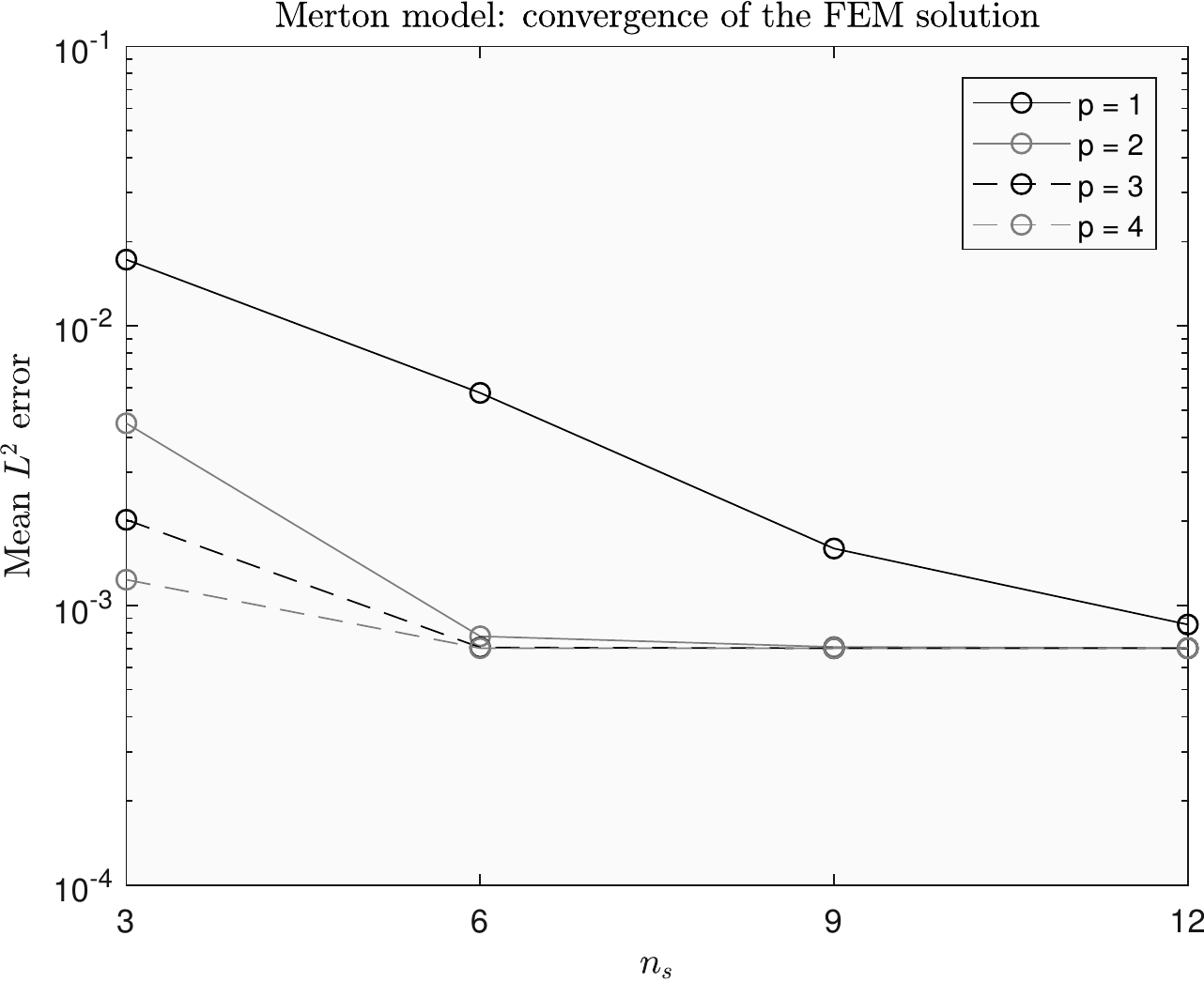}
\end{subfigure}
\begin{subfigure}{.48\textwidth}
\includegraphics[width=.99\linewidth]{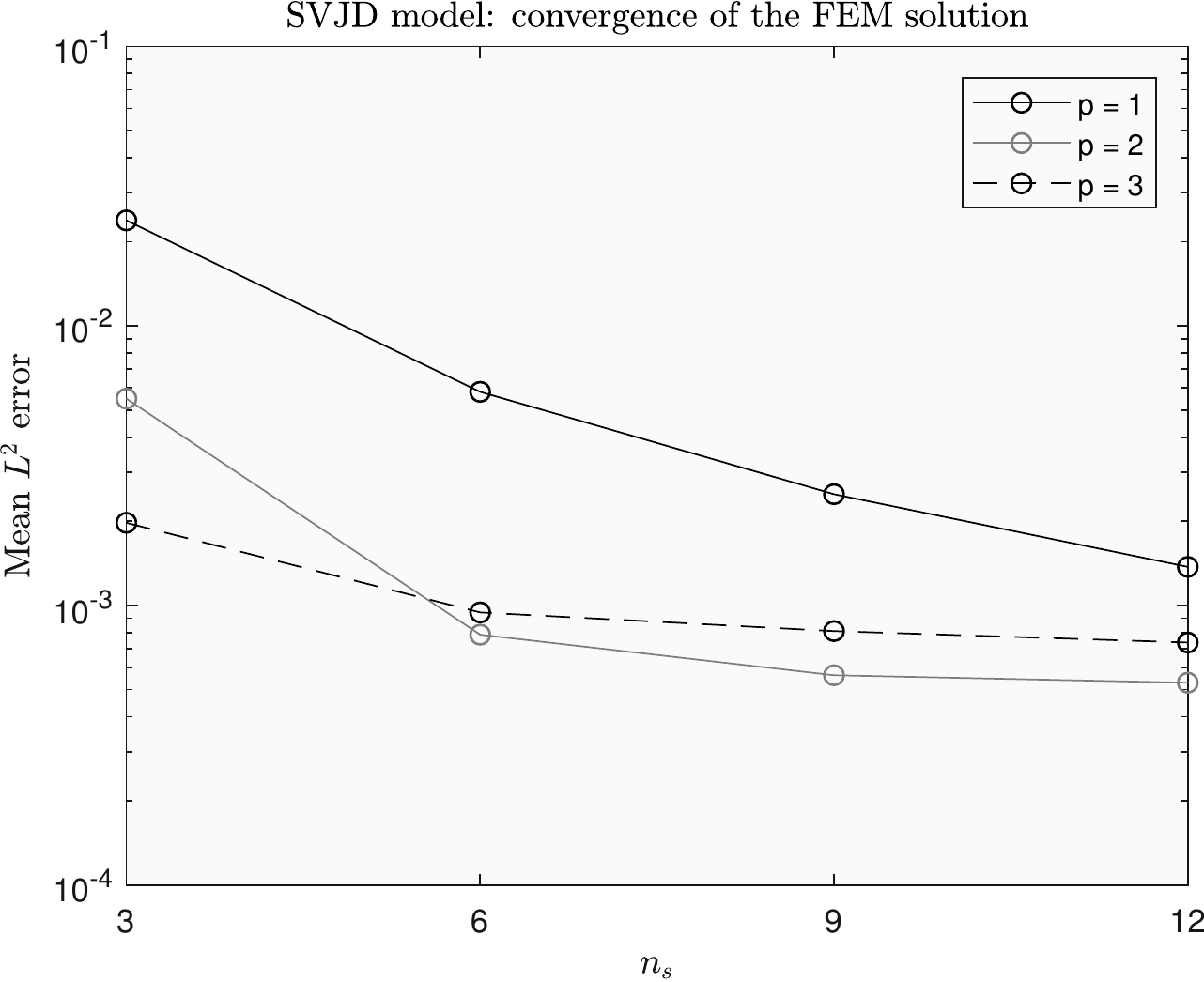}
\end{subfigure}
\caption{Results of the numerical experiments from Examples~\ref{ex:Merton3} and~\ref{ex:SVJD3}.} 
\label{fig:res2} 
\end{figure}

\begin{example}[Merton model]\label{ex:Merton3}
Let us consider the Merton model from Example \ref{ex:met1} with parameters $r = 0.048$, $\sigma=0.197$, $K=100$, $T= 1$, $\lambda = 0.19$, $\mu_J = -0.055$, $\sigma_J = 1.1$ and with parameters for the numerical solution $n_\tau=100$, $n_s= 3, 6, 9, 12$, $\bar{s} = 3 K = 300$ and with basis of the degree 1--4. Numerical solution is also compared to the semi-closed-form solution. Results are captured in Table~\ref{tab:res2} and Figure~\ref{fig:res2}. 
We can see that for a very small number of spatial steps we can get results of almost one order more accurate in the case of higher order basis functions. 
However, the accuracy of the solution seems to saturate. Such a behaviour has been observed and described in~\cite{Feng07} and can be caused by a truncation error. 
By comparison of the error table (Table~\ref{tab:res2}) and the table depicting the computation time (Table~\ref{tab:res3}), it is clear that the refinement of the spatial domain discretization leads to a better accuracy but also to higher time expenses. Thus, it is recommended to use the higher order basis for a sutiable accuracy/efficiency ratio.

\end{example}

\begin{example}[SVJD Model]\label{ex:SVJD3}
The numerical experiments for the SVJD model with parameters $r = 0.0529, \sigma = 0.51, \rho = 0.24, \kappa = 0.5, \theta = 0.19, \varepsilon = 0.003, H = 0.6, \mu_J = -0.1, \sigma_J=0.4, \lambda_J = 0.074, K = 100, T = 1, n_t = 100$ and for $n_s = n_v = 9, 18, 27, 36$ were performed. The results are captured in Table~\ref{tab:res2} and Figure~\ref{fig:res2}. We measure the mean $L^2$ error at $v=0$ with respect to the undelying price $s$. The times needed for computation are noted in Table~\ref{tab:res3}. From the comparison of different number of discretization points we can observe that sufficiently good results can be obtained already for the basis of degree $p=2$. Note, that the refinement of the spatial domain leads to higher computational demands and naturally, at higher rate than in Example~\ref{ex:Merton3}. Thus, use of the higher order basis is recommended. 
\end{example}

\begin{table}
\begin{center}
{\footnotesize
\begin{tabular}{|l|llll|lll|}
\hline
\multicolumn{8}{|c|}{Mean $L^2$ error} \\
\hline
 & \multicolumn{4}{c|}{Merton model} & \multicolumn{3}{c|}{SVJD model}  \\
$n_s$ & \multicolumn{1}{c}{$p=1$} & \multicolumn{1}{c}{$p=2$} & \multicolumn{1}{c}{$p=3$} & \multicolumn{1}{c|}{$p=4$} & \multicolumn{1}{c}{$p=1$} & \multicolumn{1}{c}{$p=2$} & \multicolumn{1}{c|}{$p=3$}\\
\hline
9   & 0.017271 &  0.005763 &  0.001599 &  0.000855  & 0.023866 & 0.005496 & 0.001977 \\
18  & 0.004486 &  0.000776 &  0.000711 &  0.000704  & 0.005812 & 0.000786 & 0.000944 \\
27  & 0.002026 &  0.000708 &  0.000703 &  0.000703  & 0.002502 & 0.000563 & 0.00081 \\
36  & 0.001237 &  0.000704 &  0.000703 &  0.000703  & 0.001374 & 0.000529 & 0.000737 \\
\hline
\end{tabular}
}
\caption{Results of the numerical experiments from Examples~\ref{ex:Merton3} and~\ref{ex:SVJD3}. The high computation time for $p=1$ and $n_s = 9$ in both model is caused by a non-suitable integration procedure }
\label{tab:res2}
\end{center}
\end{table}

\begin{table}
\begin{center}
{\footnotesize
\begin{tabular}{|l|llll|lll|}
\hline
\multicolumn{8}{|c|}{Computation time $[s]$} \\
\hline
 & \multicolumn{4}{c|}{Merton model} & \multicolumn{3}{c|}{SVJD model}  \\
$n_s$ & \multicolumn{1}{c}{$p=1$} & \multicolumn{1}{c}{$p=2$} & \multicolumn{1}{c}{$p=3$} & \multicolumn{1}{c|}{$p=4$} & \multicolumn{1}{c}{$p=1$} & \multicolumn{1}{c}{$p=2$} & \multicolumn{1}{c|}{$p=3$}\\
\hline
9  & 0.357 &  0.328 &  0.433 &  0.538  & 1.149 & 0.798 & 0.801 \\
18 & 0.362 &  0.906 &  1.231 &  1.745  & 2.421 & 2.511 & 3.065 \\
27 & 1.366 &  1.787 &  2.535 &  3.714  & 5.411 & 6.217 & 7.506 \\
36 & 2.154 &  2.971 &  4.350 &  6.262  & 11.672 & 14.511 &  15.522 \\
\hline
\end{tabular}
}
\caption{Time needed for the computation of the solutions of  Examples~\ref{ex:Merton3} and~\ref{ex:SVJD3}. Note that the elapsed times are just orientational and depend on other factors such as CPU multitasking. }
\label{tab:res3}
\end{center}
\end{table}

Both of the examples show that the time scheme makes the finite element approximation of the function less accurate compared to the static case. 
However, the accuracy of the results is still precise enough for a common use. 

\FloatBarrier
\section{Conclusion}\label{sec:conclusion}

The aim of this paper was to introduce a computational framework how to solve pricing PIDEs numerically by isogeometric analysis tools, i.e. using FEM with non-uniform rational B-spline basis functions. The framework was derived for a rather general class of stochastic volatility jump diffusion models. In detail we covered both constant volatility and stochastic volatility jump diffusion models for European style options. In particular, we solved the constant volatility \cite{Merton76} model and approximative fractional stochastic volatility model introduced by \cite{PospisilSobotka16amf} that both have a compound Poisson process in the underlying asset process with jump sizes being log-normally distributed. Other types of jumps may be easily adapted by a straightforward modification of the matrix $\mathbb{J}$ (see \eqref{eq:met85}) only. Other types of options (e.g. barrier options) can be priced just by change of the boundary conditions of the PIDE. 

The advantage of NURBS basis functions is multifold. First of all, the definition allows to use higher order functions easily. Furthermore, thank to geometric properties, NURBS can describe complicated curves and surfaces often with low number of control points. For example, non-uniform knot vectors together with the multiplicity of knots can be easily used to describe exactly non-smooth pay-off functions. Last but not least, rationality of NURBS then give us much greater flexibility (compared to the standard B-splines) in describing complicated solutions of pricing equations.

To demonstrate the advantages of NURBS, we show how B-Splines and NURBS can be fitted to smooth, non-smooth or even discontinuous functions  easily (see Examples \ref{ex:nurbs_fit_exp}-\ref{ex:nurbs_fit_digital}) with the same knot vector. We can obtain a perfect match by only repeating one knot corresponding to the point with lower continuity degree. In Section \ref{ssec:nurbs_fit}, we present results related to fitting B-Splines and NURBS to the exact formulas for both Merton and SVJD models. Fitting experiments showed that the value of the utility function (mean $L^2$ error) in the optimization problem is highly sensitive to the choice of the initial guess. Therefore, the choice of suitable NURBS weights in the FEM for studied PIDEs is still under investigation.

Standard approach to solve the pricing differential equation is to introduce the transformation $x=\ln s$, i.e. in the Black-Scholes or Heston type of models, one gets an equation with constant coefficients. In our case of a more general model \eqref{e:PIDE}, no such transform is known and hence we stay with the original variable $s$. This approach is advantageous in the sense that the pay-off functions remain untransformed and can be described precisely by the given basis functions. On the other hand, non-constant coefficients increase the complexity of numerical behaviour of the scheme and thus it makes the numerical analysis slightly more complicated. Some results show that being able to describe the initial condition of PIDE accurately does not always outweigh the disadvantages of solving a linear PIDE with non-constant non-linear (quadratic) coefficients. 

FEM with NURBS basis function is currently a hot research topic referred to as the isogeometric analysis. Although its ideas come especially from computational mechanics and computer-aided geometrical modelling, its usage in mathematical finance and econometric applications is apparent. In this paper, we presented especially the idea of the NURBS-based finite elements; however, in parabolic PIDEs, the suitable iterative scheme in time variable is also of importance. For the sake of simplicity, we considered the weighted scheme of the first order \eqref{eq:met50}, in particular, we used the fully implicit scheme in presented numerical experiments. More complex iterative schemes can of course lead to lower error and better convergence results. For example, \cite{Feng08} studied extrapolation schemes in option pricing models. 

It can be shown that for simple basis functions in finite elements in rectangular domains there exists an equivalent finite difference method. However, thanks to the above-mentioned advantages of the NURBS, the solution obtained using finite differences would hardly achieve the same properties as the NURBS-based FEM solution. In the recent paper by \cite{Hout16}, ADI finite differences schemes are used to price both European and American put options under the Merton or Bates model. Corresponding PIDEs are, therefore, of the same type as in this paper; however, authors do not provide a comparison of their method to the price obtained by a semi-closed form solution. The influence of localization of the integral term in the studied PIDEs that acts usually globally over the whole domain is not fully known for general pay-off functions and it is under investigation. 

If there exists a semi-closed form solution for studied stochastic volatility jump diffusion model, it is of course superior to any other means of numerical solution of corresponding pricing equations. However, even these formulas can have serious numerical problems as was shown, for example, by \cite{DanekPospisil17}. Let us mention at least one minor advantage of using finite elements over semi-closed formulas, namely one finite element solution give us prices of options for one strike and all maturities at once. It is worth to mention that the semi-closed formula is known only for a specific choice of the functions $p$ and $q$ in \eqref{e:PIDE}. However, the framework developed in this paper can handle the model with general $p$ and $q$. 

To support the advantages of FEM with NURBS basis functions, we compared a numerical solution of the Merton and SVJD model for European call/put option to the solution obtained by a semi-closed formula. We obtained a very good precision results with low number of discretization points.

Although convergence results for PIDEs are not completely known, they are intensively studied in communities of mathematical and numerical analysis researchers. Presented computational framework can be useful not only for theoretical researchers, but especially to practitioners who are looking for a powerful tool for pricing derivatives that involve PIDEs. Although pricing American types of options goes beyond the aims of this manuscript, since it requires a solution of the partial integro-differential variational inequalities, see for example \cite{Feng07}, it should not be difficult to modify presented framework also to these types of problems.


\section*{Acknowledgements}
Computational resources were provided by the CESNET LM2015042 and the CERIT Scientific Cloud LM2015085, provided under the programme ``Projects of Large Research, Development, and Innovations Infrastructure''.
\section*{Disclosure statement}

No potential conflict of interest was reported by the authors.

\section*{Funding}

This work was partially supported by the GACR Grant 14-11559S Analysis of Fractional Stochastic Volatility Models and their Grid Implementation. 




\end{document}